%% file: main.tex
\documentclass[12pt,fullpage]{article} 
\input{packages} 
\allsectionsfont{\bfseries\sffamily}

\usepackage{abstract}

\newtheorem{proof}{Proof}
\newtheorem{definition}{Definition}

\RestyleAlgo{ruled}

\title{\huge \bfseries\sffamily \textbf{The Problem of Dynamic Spatial Sampling and Geofence Surveillance} 
            \thanks{Corresponding Author: Marty Davidson, Email: marty.davidson@wisc.edu}
        } 
        
\author[1]{Marty Davidson} 
\affil[1]{University of Wisconsin - Madison}
\author[2]{Jason Byers} 
\affil[2]{University of Connecticut}

\newdateformat{monthyeardate}{
        \textbf{\bfseries\sffamily Current Version:} \monthname[\THEMONTH] \THEYEAR
        }
\date{} 

\begin{document}
%

\begin{singlespace}
    \maketitle 
    \begin{center}
    \begin{abstract} 

\input{2_Sections/0_Abstract}
    \end{abstract}        
    \end{center}


\end{singlespace}
\newpage

\input{2_Sections/0_Introduction}

\input{2_Sections/2_Algorithmic_Evaluations}

\input{2_Sections/2_Defining_Problem}

\input{2_Sections/3_Case_Studies/0_Introduction}
\input{2_Sections/3_Case_Studies/A_Window_Adaptive}

\input{2_Sections/3_Case_Studies/B_Focal_Adaptive}

\input{2_Sections/3_Case_Studies/C_Lambda_Adaptive}

\input{2_Sections/3_Case_Studies/D_Polygon_Extension}

\input{2_Sections/4_Simulation_Overview}

\input{2_Sections/5_Results}

\input{2_Sections/6_Choosing_K}

\input{2_Sections/7_Conclusion}


\appendix
\input{3_Appendicies/Appendix_Notation}
\input{3_Appendicies/Appendix_Scope_Analysis/0_Introduction}
\input{3_Appendicies/Appendix_Scope_Analysis/A_Poisson_Maximum_Likelihood}
\input{3_Appendicies/Appendix_Scope_Analysis/B_Cluster_Domain_Analysis}

\input{3_Appendicies/Appendix_Scope_Analysis/C_Smallest_Bounding_Sphere_Problem}

\input{3_Appendicies/Appendix_Proofs}
\input{3_Appendicies/Appendix_Window}
\input{3_Appendicies/Appendix_Focal}

\input{3_Appendicies/Appendix_Lambda}
\input{3_Appendicies/Appendix_Central}
\input{3_Appendicies/Appendix_Polygon}
\input{3_Appendicies/Appendix_Simulation}
\input{3_Appendicies/Appendix_Average_Time}

\begin{singlespace}
    \bibliographystyle{plainnat} 
    \bibliography{references}
\end{singlespace}

\end{document}

%% file: packages.tex
\usepackage{blindtext} 

\usepackage{geometry}
\usepackage{datetime} 
\providecommand{\keywords}[1]
{
  \small	
  \textbf{\textit{Keywords:}} #1
}
\usepackage{setspace}

\usepackage[numbers]{natbib}
\usepackage{appendix}

\usepackage{multicol}
\usepackage{dcolumn}
\usepackage{booktabs}
\usepackage{tabularx}
\usepackage{threeparttable}

\usepackage{amsmath}
\usepackage{amssymb}
\usepackage{amsfonts}
\usepackage{float}
\usepackage{bbm}
\usepackage{algorithm2e}
\usepackage{breqn}

\usepackage[colorlinks=TRUE,linkcolor=blue!50!gray,citecolor=red!50!black,urlcolor=blue!50!gray]{hyperref}
\usepackage{url}
\usepackage{hyperref}

\usepackage{epigraph} 
\usepackage{lscape}
\usepackage{xcolor}
\usepackage{graphicx}
\usepackage[utf8]{inputenc}
\usepackage{soul}
\usepackage{siunitx,booktabs}
\usepackage[skip=0.5\baselineskip]{caption}
\usepackage{sectsty}
\usepackage{booktabs}
\usepackage{placeins}
\usepackage{rotating}
\usepackage{authblk}
\usepackage{comment}

%% file: 2_Sections/0_Abstract.tex
Geofencing surveillance poses a dynamic spatial sampling problem. Police agencies must select a surveillance site, choose a geofence perimeter from a set of alternatives, and identify potential suspects through reverse location warrants. At the same time, warrant magistrates must impose constraints that curtail the reach of police surveillance efforts. This sampling problem emerges because agencies commonly use fixed geofence boundaries that ignore how humans move about a chosen surveillance site (i.e., pedestrian flows or traffic patterns). This further exacerbates privacy concerns and increases the risk of selective expansion where agencies extend their data collection efforts beyond the parameters outlined in their warrant. Given the Court's recent ruling in Chatrie, there is currently a need to establish a measurable process that allows magistrates to quantify and evaluate the potential impacts of a warrant proposal. In this paper, we take the first step in introducing a set of optimal radius estimators that measure how geofence perimeters adapt to their local context. Given a surveillance site and some privacy constraint, these estimators generate surveillance perimeters whose size changes with local population densities. This allows magistrates to quantify tradeoffs between local privacy intrusions with law enforcement's surveillance needs. We discuss the properties of these estimators, their underlying assumptions, and the potential consequences of using algorithms to better protect the privacy of its citizens. 

%% file: 2_Sections/0_Introduction.tex
\section*{Introduction}
During their Spring 2026 term, the US Supreme Court handed down an important fourth amendment ruling on the constitutionality of geofence warrants (see Chatrie vs. United States). Up until this point, law enforcement agencies increasingly used these warrants to identify criminal suspects by collecting historical geolocation data from third-party vendors like Google \cite{Fussell_2021, bathke2024, amster2022, chowdri2024}. Agencies would make iterative data requests from vendors based on the geographic and temporal boundaries of their geofence \cite{fan2025,berris2026}. From here, an agency whittles down and de-anonymize the  geolocation tags that happen to intersect with their geofence. 

Although the Chatrie Court struck down this iterative data collection process as unconstitutional, it left undefined the thornier questions of how an agency should construct its geofence boundary and how a warrant magistrate should constrain an agency's proposal. To protect individual privacy rights, therefore, we have to imagine a framework that empower magistrates to quantify the tradeoffs between surveillance needs and fourth amendment protections. We begin by introducing a statistical framework that can a) measure the physical properties of a geofence proposal and b) compare it against some (null) distribution of plausible perimeter alternatives. 

In this paper, we focus on the procedural dynamics between law enforcement and warrant magistrates. Specifically, how can a magistrate measure and evaluate the physical attributes of a geofence proposal? We interpret this dynamic as both a measurement problem and an optimization problem. From a measurement perspective, how can magistrates measure a proposal's attributes and compare these attributes against a potential pool of alternatives? From an optimization perspective, how can magistrates implement a privacy constraint that limits the physical properties of an agency's proposal? 

We begin our paper by reframing these dynamics as a three-stage dynamic spatial sampling (DSS) problem that focuses on how points intersect with digital geometries \cite{cheung2004}. Under this DSS framework (see Appendix \ref{appx:notation} for full notation guide), an agency selects a focal site for surveillance, a judge then specifies some quantifiable privacy constraint $k$ that incorporates local context information and minimizes bystander risk, finally, the agency chooses an "optimal" perimeter from a distribution of potential alternatives. 

Unlike Chatrie's iterative data collection process\footnote{In Chatrie, the Supreme Court scrutinized the three step sampling process the police used to collect data from Google. First, law enforcement defined a temporally bounded geofence across a specific region. Google then produced a population of anonymized geolocation tags based on the geofence specifications. Second, law enforcement reviewed this data and then asked Google to provide more detailed information on a subset of the initial population. Finally, law enforcement demanded Google hand over the de-anonymized identities for a further subset of users. This information formed the basis of the suspect pool, which included names, home addresses, email addresses, and phone numbers. The Court, however, did not rule on this issue and has left this task for the U.S. Appellate Court system to figure out on its own}, this sampling process specifies a priori the judge's privacy preferences and adapts it to some local context. In Section \ref{section:chooseK}, we outline how legal parties can use anonymized population density maps to estimate the expected number of "captured" individuals for a geofence proposal. This value reflects the anticipated number of individuals that will intersect with an agency's geofence given the population density map ($\hat{k}$). This quantifies the direct impacts a geofence proposal will have on community without de-anonymizing any individual's historical cell phone location data. 

To measure a geofence's potential impact, we introduce a set of plug-in maximum likelihood estimators that generate an optimal radius $\hat{r}$ for constructing circular or non-circular sampling regions \cite{rossi2018}. For each estimator, the estimated number of "captured" individuals depends only on i) population densities around the focal surveillance site $\lambda(a_i)$ and ii) a magistrate's privacy constraint $k$ (see Appendix 3, Proofs A and B). In this setup, auditors can mathematically optimize geofence radii to satisfy explicit privacy constraints without compromising investigative utility. In addition, they can derive a distribution of potential geofence perimeters from anonymized density maps of cell geolocation tags.

To test our plug-in estimators, we use agent-based modeling to simulate commuting pathways, which we interpret as biased random-walks. We evaluate and compare we compare their performance against what we see as the current legal default where an agency randomly selects a geofence's boundary size based on intuition. 

We conclude by simulating how magistrates can gather important insights into a geofence proposal's invasive properties. Here, we measure the selective expansion risk of the circular geofence perimeter used in Chatrie within a hypothetical neighborhood setting  \cite{amster2022}. Using only anonymized density maps, we generate a distribution of plausible boundary alternatives, which auditors can use to construct statistically meaningful sampling distributions for $\hat{r}$ (estimated radii holding $k$ fixed) and $\hat{k}$ (estimated number of intersecting units holding $r$ fixed).






%% file: 2_Sections/2_Algorithmic_Evaluations.tex
\section*{Working Example}
Throughout the paper, we will use the following hypothetical as a motivating example. Here, we outline the assumptions that operationalize what a warrant proceeding might look like moving forward. First, we assume a law enforcement agency wants to establish a circular digital perimeter of radius $r$ around a specific location $a_i$ (such as a retail shop) to sample as many individuals as possible within a finite time period (Figure 1, right graph). 

Next, to prevent over-surveillance and protect innocent bystanders, a magistrate restricts the geofence's size by imposing a privacy constraint, $k$. The constraint $k$ accounts for geographic, temporal, and qualitative factors, such as the specific locations and time frames covered, and the demographic groups impacted. We model the geofence volume as $\nu(B(a_i,r)) = f(x,k)$, where $x$ represents measurable contextual information. 

In this paper, we focus on a simplified setting where a judge sets a static privacy constraint $k$ and uses population density around the surveillance site to measure geofence performance. We seek to measure the expected number of "captured" individuals. This establishes an accept-reject boundary that each geofence can be measured against (Figure 1, left, dashed line).

In order to comply with the magistrate's preferences, the agency must optimize its proposal so that it lies directly on or below the accept-reject boundary. The red circular point in Figure 1 (left) represents the agency's proposal at time $t$; modeled here as $h(x,t) + \Delta$, where $h(x)$ represents the factors law enforcement considers and $\Delta$ represents systematic measurement error. Because individuals are non-randomly distributed across space, however, the agency's proposal plots above the judge's preference curve. In this hypothetical, the proposal over-surveils the target region and would likely be rejected by the magistrate. 

To minimize the distance between the proposal and the magistrate preference curve, the agency can either shrink the digital perimeter's radius or alter the geofence's location and timing until local conditions satisfy the constraint (e.g., approximately 2 people per square kilometer in this scenario). During the warrant process, the magistrate compares the agency's proposal against a range of plausible alternatives, which vary according to timing, location, or perimeter shape and can be mapped to our graph. 

\begin{figure}
    \centering
    \includegraphics[width=\linewidth]{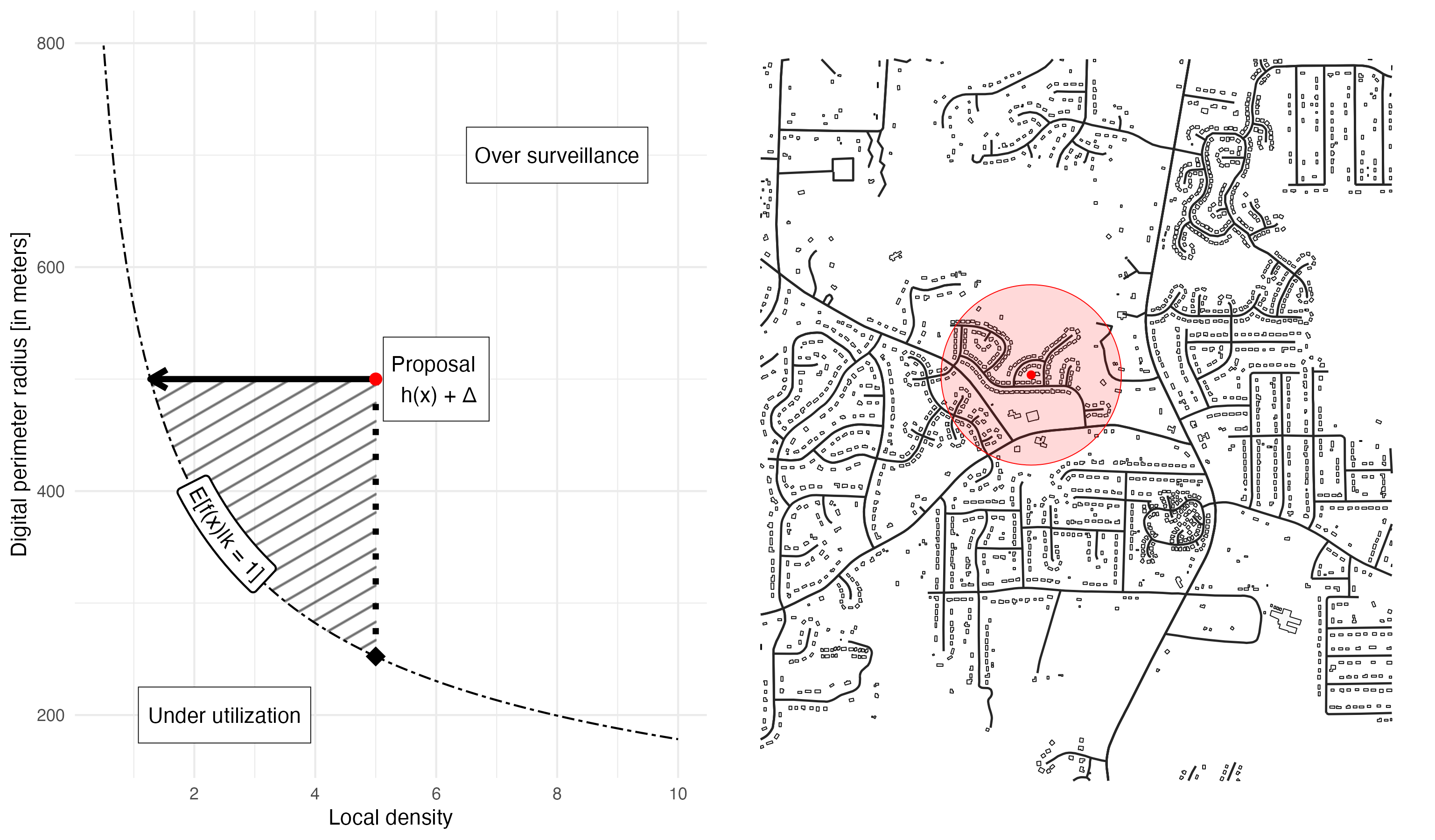}
    \caption{}
    \label{fig:figure1}
\end{figure}


%% file: 2_Sections/2_Defining_Problem.tex
\section*{Defining the Problem of Dynamic Spatial Sampling}
How can a spatial auditor construct a digital perimeter whose size depends on a spatially varying stochastic process $\eta$ and a privacy constraint $k$? This question summarizes what we define as the problem of dynamic spatial sampling. In the policing context, this problem regularly emerges when law enforcement attempts to place a geofence perimeter in a socially dynamic public setting while also being subject to an externally imposed privacy constraint $k$. Here, we frame these social patterns as a stochastic process $\eta$, which produces distinct point patterns across a jurisdiction\footnote{For now, we will ignore temporal dynamics and focus exclusively on how agencies construct spatial boundaries.}. Solving this problem requires an agency to simultaneously account for legal privacy constraints ($k$) and the highly dynamic spatial patterns of human activity ($\eta$). 

When selecting a digital perimeter, agencies tend to select fixed boundaries that might adhere to legal constraints (i.e., timing, location, and size) yet fully ignore social patterns of human activity. This matters when estimating the number of intersecting units $\hat{k}$ for a given perimeter. This can lead to unintentional selective expansions of surveillance, where an agency collects additional digital identifiers that partially extend beyond its legal mandates \cite{amster2022}. For example, we can expect selective expansion to be greater in densely populated areas, where the probability of erroneous point intersection is highest.

Assuming the agency wants a circular perimeter, building a digital perimeter involves constructing a ball shaped sampling region $B(a_i,\hat{r}_{\text{Fixed}})$ with a fixed radius $\hat{r}_{\text{Fixed}}$. This reflects the basic setup in Chatrie, where law enforcement used a circular perimeter with a radius of 150m \cite{supremeCourtChatrie}. By construction, fixed perimeters ignore available spatial information inside a jurisdictional study window: $\Omega \subset\mathbb{R}^2$. This includes, but not limited to, the size of the study window $\nu(\Omega)$, the total number of individuals $n$, the density of individuals at the focal site for surveillance $\lambda(a_i)$, and the density of individuals at arbitrarily defined locations $\lambda(\omega)$, where $\omega \in \Omega$ . 

Although simple to construct, arbitrarily defined fixed perimeters can exacerbate substantive debates over surveillance and privacy protection. Each camp maps a different set of features to $\hat{r}_{\text{Fixed}}$ in order to justify their preferred ideological positions. For example, we can expect pro-surveillance groups to prefer assigning large values to $\hat{r}_{\text{Fixed}}$ that maximize surveillance scope. This breadth provides a way to collect a sufficient number of anonymized data points that then can be whittled down to a potential suspect \cite{chatrieOpp2025}. We can expect pro-privacy groups, on the other hand, to prefer assigning small values, including zero, to minimize surveillance scope and maximize individual privacy protections \cite{chatrieOpeningBrief2026}. Further complications arise when an external authority, such as a judicial warrant officer or third-party vendor (i.e., Google), introduce additional constraints on $k$ (i.e., timing, size, location) to the discussion forum. 

Even when discussing real-world features (i.e., location and timing of the criminal event), such considerations do not explicitly condition their choices for $\hat{r}_{\text{Fixed}}$ on quantifiable information from inside the jurisdictional study window. This disconnect between the geofence perimeter construction criteria and the measurable information inside the study window generates measurement error that sets the stage for selective expansions of surveillance operations. When law enforcement arbitrarily selects a value for $r$ and fix it, the non-random distribution of human activity can undermine privacy constraints and lead to selective expansion. Fixed perimeters, therefore, will sample individuals differently depending on where the surveillance operation occurs. This inhibits the development of legal and regulatory principles that can establish a framework for using digital perimeters. 

In the remainder of this paper, we assume that the privacy constraint $k$ sufficiently encodes the preferences of the magistrate, $r$ encodes the preferences of pro-surveillance advocates, and $\lambda(\omega)$ encodes information on patterns of human activity across different locations inside the jurisdiction $\omega \in \Omega \subset  \mathbb{R}^2$. In addition, we assume that the joint distribution $\{(\lambda(\omega), k, r): \omega \in \Omega \subset  \mathbb{R}^2\}$ fully measures the debate over surveillance versus privacy tradeoffs. Finally, we assume that the best solution balances these tradeoffs. This emerges when law enforcement conditions its value for $r$ on $\lambda(\omega)$ and constrain it to a legally permissible value for $k$. 

When framed in this manner, it becomes apparent that fixed perimeters ignore the full joint distribution between all three parameters $\{(\lambda(\omega), k, r): \omega \in \Omega \subset  \mathbb{R}^2\}$. In what follows, we use maximum  likelihood estimation to derive a set of plug-in estimators that allow auditors to select a value of $r$ while conditioning on the empirical distribution of $\lambda(\omega)$. These estimators generate a set of digital perimeters whose boundary shape, size, and contours adaptively respond to different patterns of human activity. Across multiple time periods of recorded patterns of human activity (i.e., daily fluctuating commuting pathways), these optimal sampling regions return an unbiased and consistent estimate of $k$ (see Appendix 3, Proof C). Specifically, agencies can generate boundaries whose estimate values of $\hat{k}$ converge towards the privacy constraint $k$ across multiple realizations of anonymized density data.

%% file: 2_Sections/3_Case_Studies/0_Introduction.tex
\section*{Plug-In Estimators for Constructing Geofence Perimeters}
Here, we introduce our plug-in estimators for constructing geofence perimeters: \textit{window adaptive}, \textit{focal adaptive}, and \textit{lambda adaptive} with extensions to non-circular perimeters. For non-technical audiences, these estimators range from "less realistic yet statistically easy" to "more realistic yet statistically complicated." Auditors plug-in their preferred privacy constraint $k$ and empirically derived estimates of $\lambda(a_i)$, which are the observed (anonymized) densities of geolocation tags at the focal surveillance site  $a_i \in \Omega$\footnote{$\lambda(a_i) = \frac{k}{n}$ for the window adaptive plug-in estimator}. By plugging in these values, auditors can select the "optimal" point $r$ from the joint space $\{(\lambda(a_i), k, r): a_i \in \Omega \subset \mathbb{R}^2\}$. Optimal, as used here, does not imply properly identified; rather, it denotes the best estimate of $r$ given the empirical distribution of $\lambda(a_i)$. 

Auditors can use these estimators in two different contexts. The first involves constructing the geofence perimeter. Fixing the privacy constraint $k$, auditors generate a sampling distribution of radii $\hat{r}$ whose geofence perimeters capture $k$ individuals across different realizations of $\hat{\lambda(a_i)}$. This distribution can be used to draw inferences about the optimal radius $r$ (for a circular geofence perimeter) given the data. For example, is this sampling distribution normally distributed or is it skewed/multi-modal? The uncertainty of $r$ provides important insights into selective expansion risk and the capacity for an agency to collect additional data beyond their legal mandate. 

The second context assesses the performance of a geofence perimeter. Having selected an optimal radius $r$, auditors generate a sampling distribution for the estimated number of "captured" individuals $\hat{k}$. This distribution quantifies the uncertainty in estimating $\hat{k}$ across different realizations of $\hat{\lambda(a_i)}$. For example, how much does this estimated value of $\hat{k}$ differ from the specified privacy constraint $k$? This provides important insights into the stability of potential suspect pools. 

Regardless of their substantive applications, the estimated values for $\hat{r}$ and $\hat{k}$ depend only on the plug-in estimator parameters (see Appendix 3, Proof B)\footnote{Assuming that the density $\lambda(a_i)$ has no measurement error}. When populations are homogeneously distributed ($\lambda(a_i) = \lambda$) across time, their sampling distributions provide unbiased estimates of $r$ and $k$ as they converge towards a standard normal distribution (see Appendix 3, Proof C). In the following sections, we will focus on the statistical properties of these estimators. Appendices 4, 5, 6, 7, 8 contain the derivations for each estimator.  

%% file: 2_Sections/3_Case_Studies/A_Window_Adaptive.tex
\subsection*{Window Adaptive Geofence Perimeters}
Often, auditors lack comprehensive information on where individuals are located. Instead, they might only possess information on the total number of individuals across $n$ and the size of the jurisdiction $\nu(\Omega)$. When this occurs, auditors can use the window adaptive plug-in estimator $\hat{r}_{\text{Binom}}$ to construct digital perimeters. Given a jurisdiction $\Omega$ and a focal point for surveillance $a_i \in \Omega$, this estimator conditions the radius of a circular geofence perimeter on the total area of the jurisdiction $\nu(\Omega)$, the total number of persons inside that jurisdiction $n$, and the privacy constraint $k$\footnote{Where $\Omega \subset \mathbb{R}^2$ and $\nu()$ is the Lebesgue measure} (see Equation \ref{windowAdaptive}). 

We derive this estimator from the Binomial point process \cite[Section 2.2]{chiu2013}. (see Appendix 4 for technical detail). This setup assumes there is a fixed number of geolocated individuals that are uniformly distributed across a jurisdiction. Given the complexity of social life and the influence physical surroundings have on individual-level behavior \cite{lynch1960image}, this scenario rarely arises in real-world applications. These perimeters are useful, however, in contexts where patterns of human activity are assumed to vary uniformly and an agency can adequately approximate the total population of geolocated individuals. 

$\hat{r}_{Binom}$ exhibits several unique behaviors. First, the estimator scales with the size of the jurisdiction: $\nu(\Omega)$. Holding fixed the other parameters, $\hat{r}_{Binom}$ increases non-linearly at a rate of $\sqrt{\nu(\Omega)}$. Second, the estimator is inversely related to the total number of commuters $n$ inside the study window. $\hat{r}_{Binom}$ shrinks at a rate of $(\sqrt{n})^{-1}$ when the total number of commuters increase. Finally, the estimator scales with the privacy constraint $k$, where increasing the privacy constraint $k$ increases $\hat{r}_{Binom}$ at a rate of $\sqrt{k}$. 

When the total number of individuals and the size of the jurisdiction are held constant, window adaptive perimeters become fixed in their shape and size. The size of the digital perimeter will be constant regardless of the focal site for surveillance.  Unlike fixed perimeters, however, the size of window adaptive perimeters will depend entirely on the privacy constraint $k$. Conversations about the tradeoffs between surveillance and privacy are therefore reduced to conversations about the size of the privacy constraint $k$ (see Appendix 3, Proof B). 

\begin{definition}[Window Adaptive Estimator]
    \[
    \hat{r}_{\text{Binom}} = \sqrt{\nu(\Omega)\frac{k}{n \pi}} 
    \]
    \label{windowAdaptive}
\end{definition}
\vspace{-1.05cm}

%% file: 2_Sections/3_Case_Studies/B_Focal_Adaptive.tex
\subsection*{Focal Adaptive Geofence Perimeters}
Sometimes, auditors possess detailed information about patterns of human activity around their chosen focal site for surveillance. For example, an agency seeking to establish a perimeter around a public park might possess information about the average number of daily park visitors. This number might be greater (or less) than the average density of residents in the jurisdiction when residents are assumed to be uniformly distributed. When this occurs, auditors can use the focal adaptive plug-in estimator $\hat{r}_{\text{Pois}}$ to construct digital perimeters. Given a jurisdiction $\Omega$ and a fixed surveillance site $a_i \in \Omega$, this estimator conditions the radius of a circular geofence perimeter on the local density of geolocation tags at the surveillance site $\lambda(a_i)$ and the privacy constraint $k$ (see Equation \ref{focalAdaptive}). 

We derive this estimator using a homogeneous Poisson point process \cite[Section 2.3]{chiu2013} (see Appendix 5). Similar to the window adaptive setting, this plug-in estimator assumes individuals are uniformly distributed across a jurisdiction. It differs, however, by conditioning on available information at the surveillance site. We therefore swap global uniformity\footnote{The total number of individuals divided by the total size of the jurisdiction} with local uniformity\footnote{the average number of individuals at a specific location} and assume that the local density $\lambda(a_i)$ extrapolates to all regions of the study window. This extrapolation works in settings where the range of autocorrelation is relatively large, where the density estimate $\lambda(a_i)$ varies slowly when moving away from the surveillance site $a_i$.

$\hat{r}_{\text{Pois}}$ exhibits several unique behaviors. First, the estimator is inversely related to the local point density at the surveillance site: $\hat{r}_{\text{Pois}} \propto \lambda(a_i)^{-1}$. This means that as the local density of human activity increases, the size of the associated geofence perimeter will shrink at a rate of $(\sqrt{\lambda})^{-1}$. When $\lambda \rightarrow \infty$, the size of the perimeter converges towards zero, which is simply the focal site for surveillance. Finally, the estimator is proportionally related to the privacy constraint $k$, where increasing the value of $k$ increases the size of the geofence perimeter at a rate of $\sqrt{k}$. 

\begin{definition}[Focal Adaptive Estimator]
    \[
    \hat{r}_{\text{Pois}} = \sqrt{\frac{k}{\pi \lambda (a_i)}}
    \]
    \label{focalAdaptive}
\end{definition}
\vspace{-1cm}

%% file: 2_Sections/3_Case_Studies/C_Lambda_Adaptive.tex
\subsection*{Lambda Adaptive Geofence Perimeters}
The focal adaptive plug-in estimator works well when patterns of human activity vary slightly across a region. In these settings, an auditor can get away with extrapolating local uniformity across the entire study window. More often than not, however, geolocated individuals often exhibit spatial non-stationarity where they cluster and disperse as they travel along their preferred pathways. This non-stationarity greatly influences whether a geofence perimeter can satisfy its privacy constraints. If an auditor possesses information on local density estimates across all subsets of the jurisdiction $\{\lambda(\omega): \omega \in \Omega\}$, then they can use the lambda adaptive plug-in estimator $\hat{r}_{\text{Inhom}}$. 

We base this estimator geofence on inhomogeneous Poisson processes, which model the placement of unevenly distributed, yet independent, points across a study window \cite[Section 2.4]{chiu2013} (see Appendix 6). Unlike the focal plug-in, this estimator adjusts perimeter size according to local density fluctuations inside the perimeter's convex hull: $\Lambda(B(a_i, \hat{r}_{\text{Inhom}}))$. Since $\hat{k}$ is weakly monotonic when extending outwards from a focal surveillance site, we do not derive $\hat{r}_{\text{Inhom}}$ directly. Instead, we adopt a minimization scheme where we buffer the circular geofence perimeter $B(a_i, \hat{r}_{\text{Pois}})$ at variable distances between $[-r,2r]$, calculate $\hat{k}$, and minimize the squared deviations between $\hat{k}$ and $k$ at these buffered distances. 

The lambda adaptive estimator maximizes privacy protections since it attempts to minimize the error between the theoretical privacy constraint $k$ and the estimated number of captured individuals $\hat{k}$. Ironically, in order to maximize privacy protections, auditors must possess comprehensive data on patterns of human activity across a jurisdiction. Auditors can use anonymized density maps to minimize controversy. This alone, however, does not solve the issue of law enforcement having initial access to databases with revealing information. For the time being, we will set aside the normative implications of this approach and prioritize how these geofence perimeters better estimate our target parameter $k$.

%% file: 2_Sections/3_Case_Studies/D_Polygon_Extension.tex
\subsection*{Extension to Non-Circular Perimeters}
So far, we have provided estimators that only construct circular geofence perimeters. There might be settings, however, where an auditor wishes to construct a non-circular perimeter. Fortunately, we can generate non-circular geofence perimeters by applying basic trigonometric transformations to our plug-in estimators. This is possible since the relationship between $k$ and $r$ is strictly monotonic for both the window and focal adaptive plug-in estimators, assuming $\lambda(a_i)$ is constant. This strict monotonicity means that any trigonometric transformation of $B(x, \hat{r})$ will be linearly related to the privacy constraint $k$.

Let's assume a scenario where the privacy constraint is set to $k=50$. If the local density at the focal site for surveillance equals $\lambda(a_i) = 1$, then the associated radius roughly equals $\hat{r}_{\text{Pois}} = \sim 3.99$. Now, let's assume the auditor wants to restrict the geofence perimeter between the central angle range of $\frac{\pi}{2}$ ($90^o$) and $2\pi$ ($360^o$) while holding the radius constant. Since the arc length of this central angle equals $\frac{\pi}{2}$, we are effectively dividing the privacy constraint $k$ by 4. We can calculate this value by plugging the focal adaptive estimator into the formula that calculates the total area of a central angle $\frac{1}{2}r^2\theta $, where $\theta = \frac{\pi}{2}$ and $\lambda(a_i) = 1$ (see Equation \ref{eq:radianExample}). 

\begin{equation}
    \frac{1}{2}r^2\theta = \frac{\theta}{2}\bigg (\sqrt{\frac{k}{\pi \lambda (a_i)}}\bigg )^2 = \frac{\theta k}{2\pi\lambda (a_i)} = \frac{\pi k}{4\pi \lambda(a_i)} = \frac{k}{4 \lambda(a_i)} = \frac{50}{4 \cdot1} = 12.5
    \label{eq:radianExample}
\end{equation}

If we can generate new privacy constraints by geometrically transforming the areas of the circular sampling regions, then we can also perform the inverse task. Specifically, we can generate new plug-in estimators that incorporate information about the trigonometric transformations themselves whilst preserving the privacy constraint $k$. In Equation \ref{eq:central}, we introduce the central angle adaptive plug-in estimator, which estimates the radius of a central angle sector whose arc length equals $\theta$. In this setup, the radius of the central angle converges to $\hat{r}_{\text{Pois}}$ as the central angle increases. When the central angle equals $2\pi$, $\hat{r}_{\theta}$ will equal $\hat{r}_{\text{Pois}}$.

\begin{definition}[Central Angle Adaptive Estimator]
    \begin{equation}
        \hat{r}_{\theta} = \sqrt{\frac{2k}{\lambda (a_i)\theta}}
    \end{equation}
    \label{eq:central}
\end{definition}
\vspace{-0.85cm}

The same principle applies when transforming circular sampling regions into regular $p$-sided polygons. We estimate these non-circular regions using the polygon adaptive plug-in estimator (see Equation \ref{eq:polygon}). For this to work, auditors must choose the number of sides $p$ for their polygon of choice. Next, they must calculate the optimal circumradius of the circle that circumscribes this $p$-sided polygon given $k$ and $\lambda(a_i)$. Finally, they can use Equation \ref{eq:polygon} to calculate the vertices for this inscribed $p$-sided polygon. Here, the radius of the circumscribing circle is set to the plug-in estimator $\hat{r}_{\text{Polygon}}$, $p$ is the number of sides for the polygon, $i$ is the number of the vertex, $(x,y)$ are the coordinates for the focal surveillance site $a_i$, and $\gamma$ is a rotation shift about the focal point. 

\begin{definition}[Polygon Adaptive Estimator]
    \begin{equation}
        \hat{r}_{\text{Polygon}} = \sqrt{2\cdot\frac{k}{\lambda(a_i)\cdot p \cdot \sin(\frac{2\pi}{p})}} 
    \end{equation}
    \label{eq:polygon}
\end{definition}
\vspace{-0.85cm}

\begin{definition}[Function for Finding Vertices of Optimal Polygon Region]
    \begin{equation}
        V_p(i) =\{x_i, y_i\} =\bigg \{{\cos}\bigg(\frac{2\pi i}{p} + \gamma\bigg)\hat{r}_{\text{Polygon}} + x, {\sin}\bigg(\frac{2\pi i}{p} + \gamma\bigg)\hat{r}_{\text{Polygon}} + y \bigg \} \; \; \; \; \forall \; i \in [0,p]
    \end{equation}

\end{definition}

%% file: 2_Sections/4_Simulation_Overview.tex
\section*{Agent-Based Model: Simulating Commuting Pathways}
In this section, we use a multi-agent model to evaluate how well each plug-in estimator quantifies tradeoffs between surveillance and privacy. Specifically, we use agent-based random walks to simulate geographic patterns of human activity (see Appendix 9 for technical details). During their commute, each agent produces a set of cell phone geolocation tags identifying their current location and previous steps. To introduce a level of realism, we allow our agents to interact (repulsion/clustering) with each other and fixed features inside the simulated study window. In addition, agents can have randomized starting positions or they can vary spatially according to a Gaussian random field. Together, we believe this generates the necessary variation we need for drawing inferences about perimeter performance. 

For each simulation round, we randomly generate a rectangular study window (of variable size), randomly select a value for the number of agents $n$, and randomly select a privacy constraint $k$ that is less than or equal to the total number of agents. Next, we randomly select 30 focal sites for surveillance $a_i$. Each site receives a set of geofence perimeters for fixed, window, focal, and lambda adaptive estimators. We then adjusted the adaptive boundaries according to the privacy constraint $k$, the total number of agents $n$, and the local density of agents at each surveillance site $\lambda(a_i)$ across each time (step) period. We treat these perimeter-site units as our primary unit for analysis. 

To simulate fixed boundaries, we randomly select a value between the range $[0.5, h]$, where $h$ represents the hypotenuse of the rectangular study window. We recognize that real world scenarios take a more sophisticated approach to generating fixed boundaries. These approaches, however, rarely communicate what considerations were made beyond choosing a surveillance site and a radius of interest. For Chatrie (2026), law enforcement considered 150 m (492 ft) to be sufficient with no additional justifications. For the purpose of this simulation, however, we simply seek to evaluate whether our adaptive boundaries perform better than a null setting where agencies randomly select a boundary with total disregard for any privacy constraint. 

One cause for concern involves updating boundary sizes for each time period. This act of updating could be driving performance as error terms in each time period cancel each other out. Here, we want to guarantee that performance is driven by the estimator itself. To draw fair comparisons with fixed perimeters, therefore, we include two additional geofences derived from the focal and lambda estimators. We fix each boundary to their optimal size at time (step) zero based on (density of agent's starting positions) and hold them constant across each time (step) period. If these fixed boundaries also show improved performance, then there is greater evidence that each estimator actually contributes a new signal beyond averaging out white noise. 

\begin{figure}
    \centering
    \includegraphics[height = 3.5in, keepaspectratio, width = \linewidth]{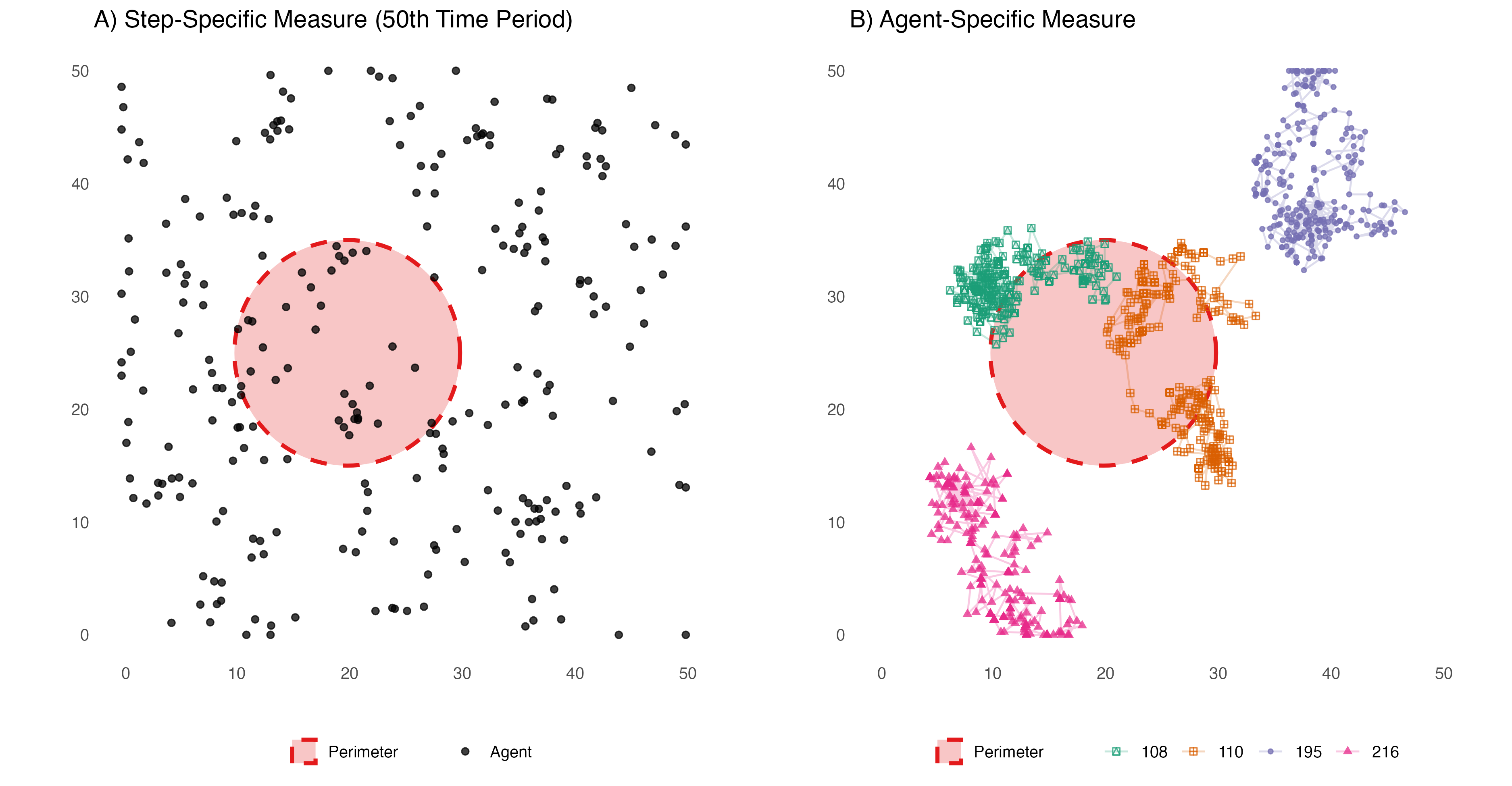}
    \caption{Outcome Measures for Agent-Based Simulation}
    \label{fig:outcomeMeasures}
\end{figure}

After constructing the perimeters, we allow our agents to conduct their random walks according to the simulation round's randomly selected parameters. As each agent walks and produces geolocation tags, we count the total number of intersecting tags $\hat{k}$ for each geofence perimeter at each focal site. We use these raw scores to generate two outcome measures.  First, we generate a step specific measure where we count the number of intersecting agents for each step round. The unit of analysis for this measure is the number of agents. Second, we generate an agent specific measure where we count how many steps per agent intersected with each perimeter. The unit of analysis for this measure is the number of geolocation tags per agent. Figure \ref{fig:outcomeMeasures} visualizes these outcome measures using a hypothetical example where 263 agents conduct a random walk of 500 steps. Here, the left figure visualizes the step-specific measure while the right figure visualizes the agent-specific measure for four agents.

For our agent-specific measure, agents experience a higher degree of surveillance when they spend more time inside a geofence perimeter. This value is bounded between [0,1], where agents are under surveillance 100 percent of the time if all their steps were inside the geofence and 0 when all their steps were outside. If an agent randomly distributed their geolocation tags across all steps $s$, then each tag has a probability of $p = \nu(B)/\nu(\Omega)$ of landing inside the geofence $B$, where $\nu(B)$ is the size of the geofence and $\nu(\Omega)$ is the size of the study window. Across all time (step) periods, the expected number of steps inside the geofence would equal $s \cdot p$. 

With this setup in mind, we measure performance using two different outcomes. First, using our step-specific measure, we examine the mean absolute deviation between $\hat{k}$ and $k$ for each perimeter type. For a single simulation round, we average the deviations for each perimeter across all surveillance sites. Second, using our agent-specific measure, we examine the mean absolute deviation between an agent's observed time under surveillance and their expected time under surveillance if agents were randomly distributed across the study window across all step periods\footnote{We assume the geofence perimeter captures exactly $k$ agents with no measurement error across all step rounds. This expected value is simply the proportion of agents under surveillance $k/n$ (see Appendix 9)}. 

For both outcomes, smaller values imply better performance. For our first measure, better performing perimeters provide better estimates of $\hat{k}$ and deviate less from $k$. For our second measure, better performing perimeters can fairly balance surveillance with privacy. This implies that agents are being surveilled at the expected rate given the privacy constraint $k$. Poor performance, on the other hand, implies that the perimeter either over- or under- surveils agents given the privacy constraint $k$.

%% file: 2_Sections/5_Results.tex
\section*{Results}
We ran 300 simulation rounds where we varied the size of the study window, the total number of agents, and random walk parameters. For each round, we randomly sampled 30 different surveillance sites across the study window. This generated 9,000 data points for simulation round / surveillance sites. For each round/site, we analyzed six geofence perimeters. In total, we analyzed 54,000 data points for simulation round / surveillance site / perimeter type. We used random forest estimation to analyze results  as it provides a flexible way to model the interactions between our simulation parameters without having to specify them directly\footnote{Given the high correlation between simulation rounds, model tuning parameters for the random forest provided non informative results. The optimal parameter for mtry, for example, was the total number of features included in the model. This effectively removed the random component of the algorithm. Instead, we used the default parameters from the ranger package in R \cite{ranger2017}: mtry = 5, min.node.size = 25, num.trees = 500. The exception is the min.node.size, which we increased from 5 observation to 25.} \cite{breiman2001}. In this section, we analyze how our performance measures correlate with the number of agents inside the window $n$ and the privacy constraint $k$.  This setup allows us to evaluate how well each perimeter performs when varying population totals and privacy constraints. To generate standard errors, we bootstrap our random forest estimates \cite{sexton2009}. 

\subsection*{Deviation from Privacy Constraint $k$}
How much does each perimeter type deviate from their randomly chosen privacy constraint $k$? Table \ref{table:deviance} summarizes partial dependence estimates for each perimeter along with bootstrapped standard errors and confidence intervals. We further breakdown the estimates by agent starting position: random locations or gaussian random field. According to this table, on average, the three adaptive perimeters - window, focal, and lambda - deviated less from the privacy constraint $k$ than the fixed perimeter. This includes the two fixed perimeters for the focal and lambda adaptive estimators. Window adaptive perimeters saw approximately a 64 percent reduction in its mean absolute deviation. Focal adaptive perimeters saw approximately a 30 percent reduction. Finally, lambda adaptive perimeters saw approximately a 94 percent reduction. These reductions were similar regardless of whether agents had random starting positions or not. 


\begin{table}[hbt!]
\begin{center}
\begin{threeparttable}%
\begin{tabular}{llllll}
\toprule
Perimeter Type & Random Start & Mean & SE & q0.025 & q0.975 \\ 
\toprule
Fixed & No & 303.6 & 8.04 & 288.8 & 319.0 \\ 
 & Yes & 294.3 & 7.75 & 280.8 & 309.3 \\ 
 \midrule
Window & No & 135.2 & 6.42 & 123.7 & 148.5 \\ 
 & Yes & 133.7 & 6.07 & 124.2 & 145.2 \\ 
 \midrule
Focal & No & 212.6 & 8.80 & 196.6 & 228.9 \\ 
 & Yes & 217.1 & 8.80 & 201.1 & 232.5 \\ 
 \midrule
Focal (Fixed) & No & 215.5 & 9.01 & 198.2 & 231.3 \\ 
 & Yes & 219.5 & 8.96 & 201.5 & 235.0 \\ 
 \midrule
Lambda & No &  16.5 & 1.74 &  13.5 &  20.1 \\ 
 & Yes &  15.5 & 1.75 &  12.9 &  19.2 \\ 
 \midrule
Lambda (Fixed) & No &  18.8 & 1.80 &  15.8 &  22.6 \\ 
 & Yes &  17.5 & 1.75 &  14.9 &  21.2 \\ 
\bottomrule
\end{tabular} 
\end{threeparttable}
\end{center}
\caption{}
\label{table:deviance}
\end{table}

\begin{figure}
    \centering
    \includegraphics[height = 3.5in, keepaspectratio]{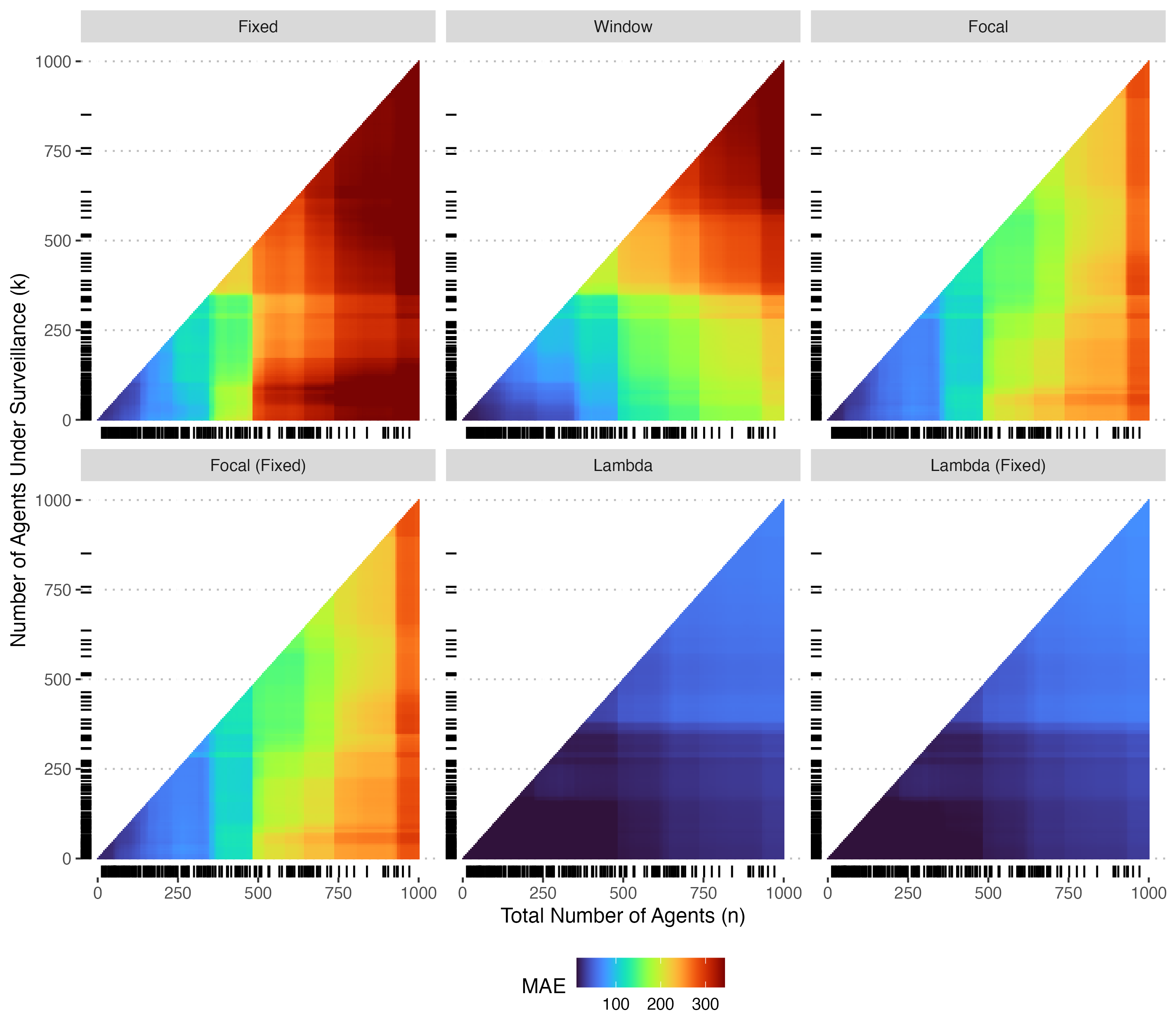}
    \caption{Bivariate Partial Dependence Plot (Deviation)}
    \label{fig:bivariateDeviation}
\end{figure}

\begin{figure}
    \centering
    \includegraphics[height = 3.5in, width = \textwidth, keepaspectratio]{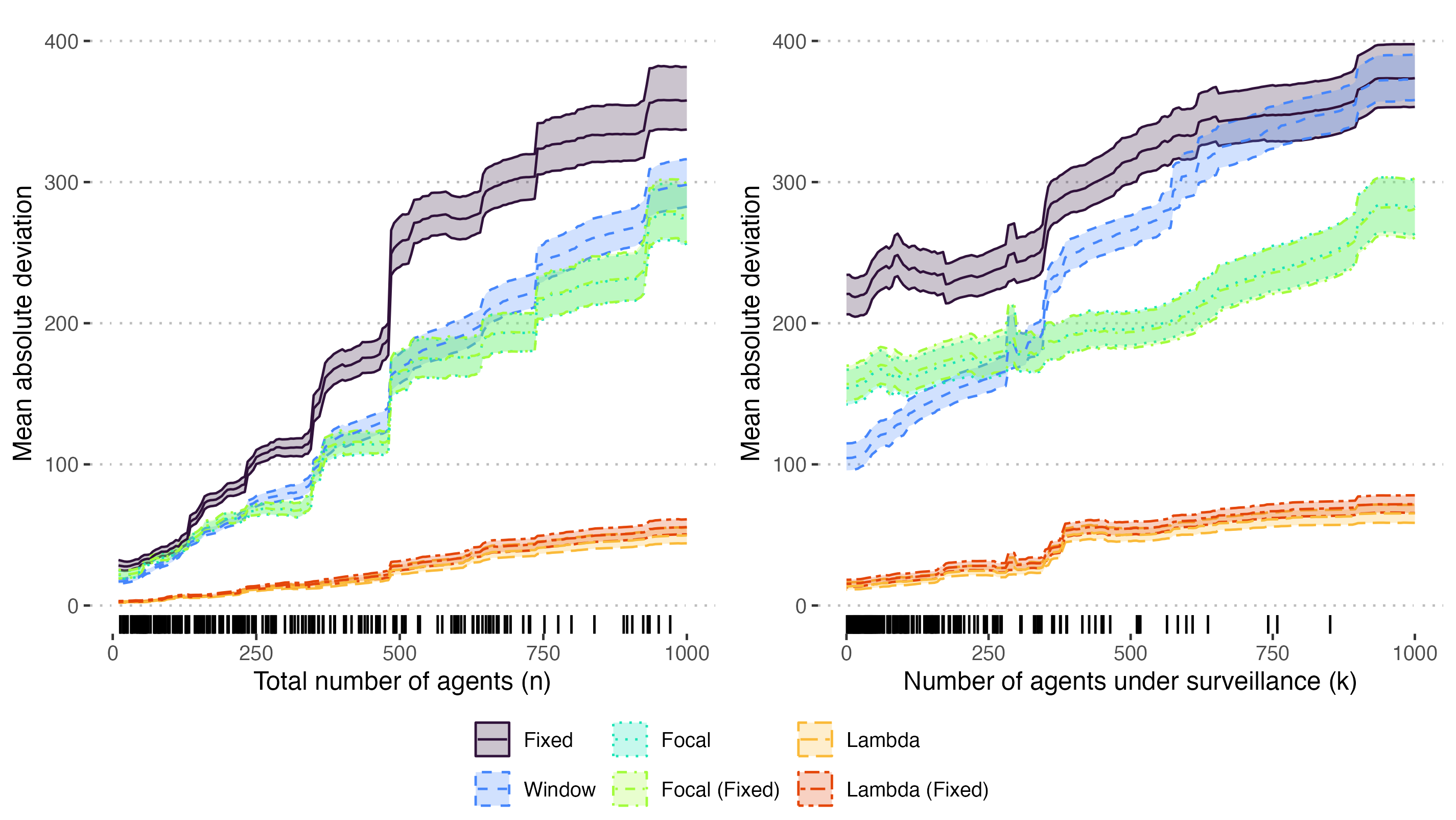}
    \caption{Marginal Partial Dependence Plot (Deviation)}
    \label{fig:marginalDeviation}
\end{figure}

The partial dependence plot in Figure \ref{fig:bivariateDeviation} evaluates how the bivariate relationship between the total number of agents ($n$) and the privacy constraint ($k$) influences the mean absolute deviation across each perimeter type. Figure \ref{fig:marginalDeviation} illustrates the marginal partial dependence plots between the total number of agents and the privacy constraint. The mean absolute deviation increases as the number of agents and privacy constraint increase. Each perimeter, however, increases at different rates. 

Here, the fixed, window, and focal (adaptive and fixed) perimeters perform similarly when the total number of agents is small. This changes as the number of agents increase. The lambda (adaptive and fixed) perimeters exhibit the best performance overall as their rate of increase remains relatively small when compared to the other perimeter types.  

Overall, the lambda adaptive (adaptive and fixed) perimeters perform best. Window and focal (adaptive and fixed) also perform better than fixed perimeters. As the privacy constraint $k$ increases, the focal (adaptive and fixed) perimeters maintain their performance gains relative to fixed perimeters. Window perimeters, however, do not. Its relative performance gains begin to fade as the privacy constraint increase. 

\subsection*{Average Time Under Surveillance}
Appendix \ref{appx:average_time} shows results for the average time under surveillance for each perimeter type. 

%% file: 2_Sections/6_Choosing_K.tex
\section*{Measuring Risk of Selective Expansions for Surveillance}
\label{section:chooseK}
\begin{figure}
    \centering
    \includegraphics[width=\linewidth]{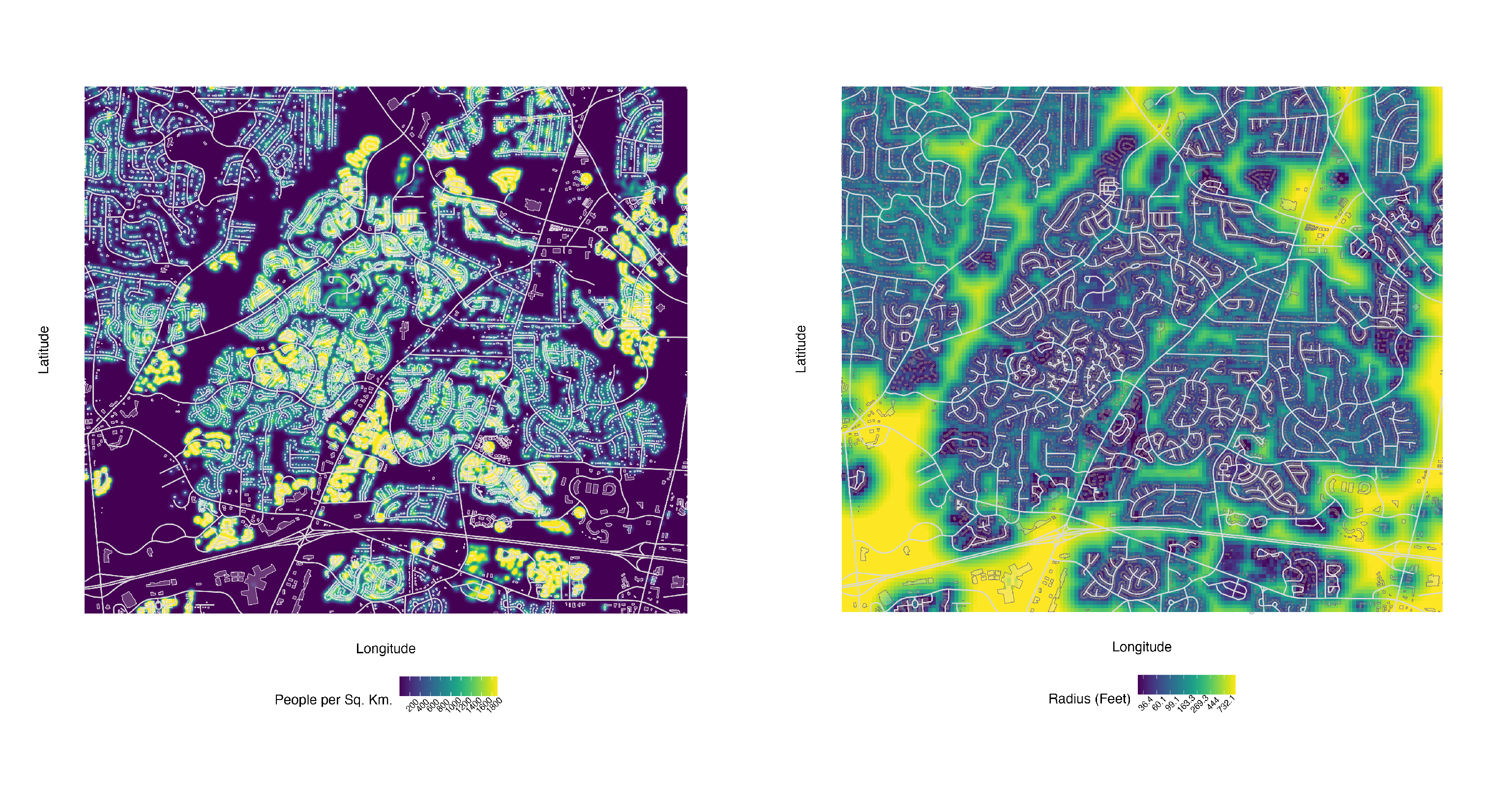}
    \caption{Residential address densities from voter file (left) and optimal radii (right)}
    \label{fig:AppliedDensity}
\end{figure}

To measure the (un-) reasonableness of a perimeter, researchers can apply the plug-in estimators to publicly available datasets, such as public voter files, to generate null distributions of plausible geofence alternatives. Given the data type and assumptions, magistrates can assess the impact that an agency's proposal might have on a community. Here, we draw on publicly available data from the North Carolina voter database and analyze the distribution of optimal radii based on the population densities for a randomly chosen neighborhood (see Figure \ref{fig:AppliedDensity})\footnote{We assume that the distribution of commuters in this neighborhood is directly proportional to the density of residential addresses inside the voter file. In other words, people work close to where they live and the population density does not change drastically over the day.}. 


Using the population densities, we generate the distribution of optimal radii through the following steps. First, we place a square grid over the density map. Next, we assume the centroids for each grid cell are different focal sites for surveillance. Finally, using each centroid, we use the lambda adaptive estimator to estimate the optimal radius at each centroid where we set the privacy constraint $k$ equal to one. The left graphic in Figure \ref{fig:AppliedDensity} visualizes these optimal radii across our neighborhood. Depending on the location, these values range from less than 30 feet to over 700 feet; however, most values are less than 100 feet. We calculate the population density by dividing $\hat{k} = 1$ by the total area of the optimal geofences $\pi r^2$.

\begin{figure}
    \centering
    \includegraphics[height = 3.5in, keepaspectratio]{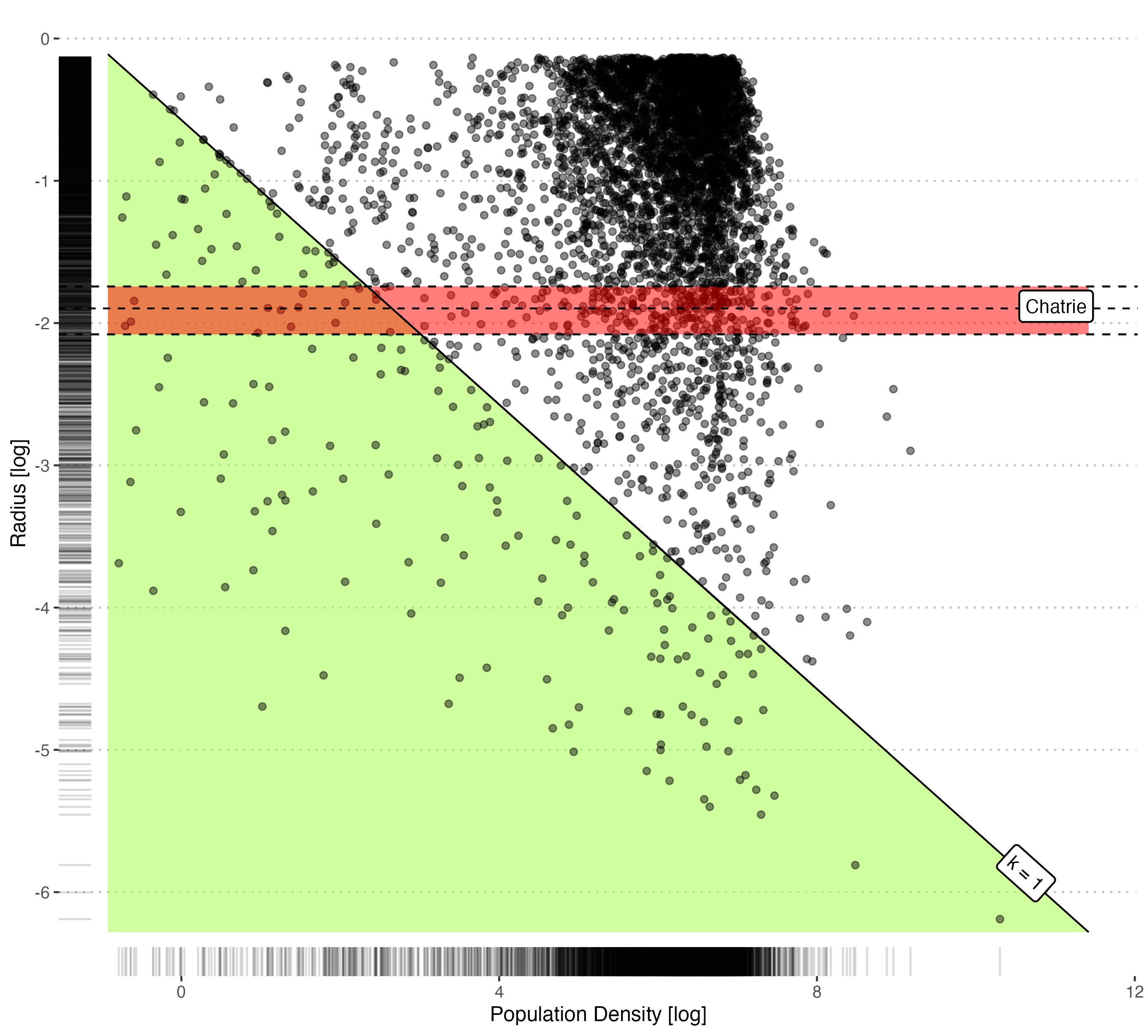}
    \caption{Scatter Plot of Potential Geofences}
    \label{fig:selectiveExpansion}
\end{figure}

To evaluate whether an agency's proposal is "reasonable", we generate a "null" distribution of randomly chosen focal sites. We then randomly assign each site a radius that we uniformly sample from the distribution of optimal radii. We then construct a set of geofences. For each, we  measure the expected number of "captured" individuals $\hat{k}$ as well as their population density by dividing $\hat{k}$ by the size of the geofence $\pi r^2$. In Figure \ref{fig:selectiveExpansion}, we plot the bi-variate relationship between the randomly assigned radii (log) and the estimated population densities (log). This chart provides a way to contextualize digital perimeter proposals across different hypothetical scenarios. Each point represents a geofence from the "null" distribution. The line represents the optimal geofences whose total areas return one individual. The green shaded region represents the optimal zone where each geofence returns less than or equal to one individual. 

According to Figure \ref{fig:selectiveExpansion}, if law enforcement were to randomly select a surveillance site in our hypothetical neighborhood and randomly assign it a radius between 6 feet and 2900 feet, then the probability that the geofence captures one individual (or less) is approximately 10 percent. Approximately 90 percent of the time, law enforcement will select a geofence that over-surveils local residents. Even in a scenario where law enforcement attempts to adhere to the privacy constraint $k$, the uneven population densities unintentionally allow the agency to selectively expand the scope of their surveillance. 

In Chatrie, law enforcement used a circular geofence with a radius of 150 meters (492.1 feet) to conduct their surveillance operation. Without proper context, it becomes difficult (from a technical perspective) to judge whether the chosen boundary proves reasonable. If an agency were to implement a similar digital perimeter in our hypothetical neighborhood, then we can plot the location of the proposal and calculate a probability score that law enforcement will over surveil local residents. Here, the middle dashed line represents the radius used in the Chatrie case. 

We sample hypothetical perimeters in a local neighborhood around this cutoff, or between 125m (410ft) and 175m (574ft). Within this region, there is approximately a 17.3 percent chance that the geofence captures one individual or less. For our hypothetical neighborhood, the geofence used in Chatrie would only be appropriate within the dark blue regions of Figure \ref{fig:selectiveExpansion}. These dark blue regions represent local parks and other sparsely populated regions. If an agency were wanting to conduct a surveillance operation in the residential or commercial regions, then they would have to employ a smaller digital perimeter. 

This example is only meant to illustrate how analysts can apply the plug-in estimators to real world applications. Our use of a static density map reflects the reality that most analysts cannot access spatio-temporal datasets that accurately measure patterns of human activity. Analysts can adopt more sophisticated datasets and approaches that are uniquely applicable to their legal scenario. Nonetheless, this exercise provides a way to understand how estimates from these plug-in estimators help contextualize geofence proposals. Specifically, they provide a way to estimate selective expansion capabilities, or the probability that law enforcement will over-surveil local residents given uneven population densities.

%% file: 2_Sections/7_Conclusion.tex
\section*{Conclusion}
In dealing with the issue of digital surveillance, technocratic expertise can complement these substantive legal debates. As it currently stands, however, very little statistical guidance exists detailing how these digital boundaries should be constructed, let alone regulated \cite{auteri2023}. This generates a principal-agent problem where judicial warrant officers mandate certain legal requirements yet struggle to generate objective performance measures that accurately assess how well law enforcement behavior conforms to these standards \cite{sappington1991}. As we demonstrated, the plug-in estimators we proposed here provides a way for non-technical auditors to construct optimal sampling regions without having to understand stochastic geometry. 

Several easy-to implement steps are required when applying our plug-in estimators to real world applications. First, auditors must select a surveillance focal site $a_i$. Second, auditors must estimate the local point density at $a_i$: $\lambda (a_i)$. This requires drawing on datasets with appropriate spatio-temporal variability into patterns of human activity. Third, an outside authority must specify their privacy constraint $k$, or the desired number of "captured" individuals, for the surveillance operation. Fourth, an auditor plugs in the lambda estimate - $\lambda(a_i)$ - and the privacy constraint - $k$ - into the plug-in estimator to generate the radius estimate - $\hat{r}$. Finally, auditors use this radius estimate - $\hat{r}$ - to construct the optimal circular geofence perimeter - $B(a_i, \hat{r})$. Additional steps are then taken for constructing non-circular perimeters or for settings where local densities vary spatio-temporally.

The approach outlined in this paper provides a way to have conversations about the reasonableness of geofence proposals. By contextualizing the observed proposal with hypothetical proposals, analysts can measure selective expansion risk and assign a probability score measuring the likelihood that the observed proposal will over-surveil local residents. It is important to note that a statistical framework alone cannot resolve all issues related to selective expansion nor Fourth amendment rights more generally. Ultimately, the communities that are being surveilled need to have input into discussion forums about community surveillance. Nonetheless, the approach used here provides an additional tool that can hopefully empower others.

%% file: 3_Appendicies/Appendix_Notation.tex
\section{Appendix 1: Notation Guide}
\label{appx:notation}
In this paper, we will use the following notation. For a continuous point process $\eta$, we will refer to its collection of points as $\{a_i\}_{i =1}^{\infty}$, where $a_i$ represents the location of the $i$-th focal site in $\mathbb{R}^d$, $a_i \in \mathbb{R}^d$.  If $a_i$ is a point that belongs to $\eta$, then we will use the notation $P(\eta(\{a_i\}) > 0)$ to denote the presence and inclusion of that specific point. We will also refer to $\eta(\cdot)$ as a random counting measure that returns 1 if a point inside $\eta$ and 0 otherwise. For each focal site,  we will use \( \lambda(a) \) when referring to the local density of human activity. Definition \ref{def:intensityFunction} defines this intensity function, where \( da \) is an infinitesimal region around \( a \), and \( \mathbb{E}[\eta(da)] \) is the expected number of points in \( da \).

\begin{definition}[Intensity Function]
    \begin{equation}
        \lambda(a) = \lim_{|da| \to 0} \frac{\mathbb{E}[\eta(da)]}{|da|}
    \end{equation}
    \label{def:intensityFunction}
\end{definition}
\vspace{-0.85cm}

\noindent We will use \( A \subseteq \mathbb{R}^d \) when referring to a borel subset, or subregion of the study window $\mathbb{R}^d$. Let \(\nu(A)\) be the Lebesgue measure that returns the area or volume of $A$. We will use \( \mathbb{E}[\eta(A)] \) or \( \Lambda(A) \) when referring to the intensity measure. This function gives the expected number of points in the measurable region $A$. We will use \( \mathbb{P}[\eta(A) = k] \) when referring to the probability that there are exactly \( k \) points in the region \( A \). Definition \ref{def:intensityMeasure} defines this intensity measure, where $\lambda(a)$ is the intensity function at location $a$ and \( da \) is an infinitesimal region around \( a \). 

\begin{definition}[Random Point Process Intensity Measure]
    \begin{equation}
        \Lambda(A) = \int_A \lambda(a) \, da
    \end{equation}
    \label{def:intensityMeasure}
\end{definition}
\vspace{-0.95cm}

\noindent For a stationary and isotropic point process, the expected number of points in $A$ is proportional to its area. Specifically, $\lambda(a)$ equals the constant $\lambda$,which can be integrated out of the integral term in Definition \ref{def:intensityMeasure} and placed in front. When a point process exhibits non-stationarity, the intensity function $\lambda(a)$ depends on the location $a \in A \subseteq \mathbb{R}^d$.

We use \( B(a_i, r) \) to denote a circular sampling region inside the study window. Here, the radius of the circle extends $r$ units from the focal point $a_i \in \mathbb{R}^d$. Definition \ref{def:BallRegion} formally defines this ball region as the collection of all Euclidean distances between $a_i$ and $y$ that are less than the radius of the circle $r$. We initially define each geofence perimeter, including non-circular regions, with a ball region. 

\begin{definition}[Circular Sampling Region]
    \begin{equation}
        B(a_i, r) = \{y \in \mathbb{R}^d : \|x - y\| \leq r \}
    \end{equation}
    \label{def:BallRegion}
\end{definition}

%% file: 3_Appendicies/Appendix_Scope_Analysis/0_Introduction.tex
\section*{Appendix 2: Discussion on Scope of Analysis}
The scope of analysis for this paper is relatively straight forward: \textit{how can spatial auditors construct sampling regions - geofence perimeters - that satisfy an optimization constraint - desired number of "captured" individuals?} Three additional research strains, however, share overlapping thematic elements that, on a superficial level, can easily distort this paper's contributions. These topics include \textit{maximum likelihood estimation for point processes}, \textit{cluster domain analyses on point interactions}, and the \textit{smallest bounding sphere problem}. In the next few paragraphs, we clearly explain the key differences that distinguish this paper's contributions from these previous research lines.

%% file: 3_Appendicies/Appendix_Scope_Analysis/A_Poisson_Maximum_Likelihood.tex
\paragraph*{Maximum Palm Likelihood Estimators for Point Processes}
Previous works on maximum likelihood estimation for cluster point processes have proposed several plug-in estimators that allow researchers to draw inferences about the cluster process' parameters, which includes the radius of interaction \cite{tanaka2008, moller2007,ogata1991}. These works tend to use a process' second-order properties (point interactions) to characterize the Palm distribution of the cluster process. \citeauthor{tanaka2008} (2008), for example, propose an approximation method for estimating Neyman-Scott parameters that re-interprets the Palm intensity function of the original cluster process as the intensity function for a difference process\footnote{A process that consists of all differences $x - y$ of points, where $D = \{x-y:x \text{ and } y \in N \text{ with } x \neq y\}$}. They set the (first-order) intensity function of this difference process equal to the Palm intensity function of the original process and use the maximum Palm likelihood estimator to estimate the Neyman-Scott parameters \cite[p.47]{tanaka2008}. In this paper, we largely ignore a process' second-order properties. We assume that point interactions between the focal surveillance site $a_i$ and geolocation tags are effectively zero\footnote{This assumption holds in settings where law enforcement agencies randomly sample their focal sites for surveillance. }. Geolocation tags can exhibit interpoint correlations, which would be measured through $\lambda(a_i)$; however, the plug-in estimators do not directly use information about the second-order properties between geolocation tags.




%% file: 3_Appendicies/Appendix_Scope_Analysis/B_Cluster_Domain_Analysis.tex
\paragraph*{Cluster Domain Analysis}
Previous works on cluster domain analysis have proposed plug-in estimators that provide a way to describe point interactions \cite{daley1971, krickeberg1973, ripley1976, kiskowski2009}. The most prominent of which is the the Ripley's K-function (Equation \ref{eq:ripleyK}) and its linear transformation (Equation \ref{eq:linearK}). Ripley's K-function, $\hat{K(r)}$, is typically used to compare a given point distribution with a random distribution under complete spatial randomness (CSR) \cite{kiskowski2009}. Here, $\hat{E(r)}$ sums the total number of points that intersect with a ball region of radius $r$ across all test points. This equals the weighted sum of points less than or equal to distance $r$ from each test point, where $w_{ij}$ are edge correction weights [cite].

\begin{equation}
    \hat{K(r)} = \frac{\hat{E(r)}}{\lambda} = \frac{\nu(\Omega)}{n^2} \sum_{i=1}^n \sum_{j \neq i}w_{ij} I(d_{ij} \leq r)
    \label{eq:ripleyK}
\end{equation}

The sampling distribution "of $\hat{K}$ is not known exactly, even for a Poisson process" \cite[p.260]{ripley1976}. Researchers can, however, indirectly analyze the sampling distributions of the centered, $\hat{H}(r)$ and linear, $\hat{L}(r)$, transformations of $\hat{K(r)}$ via convergence tests \cite{saunders1977, ripley1976}. \citeauthor[1977]{saunders1977}, for example, demonstrate how $\hat{K(r)}$ converges weakly to a homogeneous Poisson process whose constant rate, $\lambda$, is proportional to the area of any $d$-dimensional hypersphere\footnote{$\lambda r^d$, where $r$ is the radius of a $d$-dimensional hypersphere}. Asymptotic results provide an approximate sampling distribution for $\hat{H}(r)$ and $\hat{L}(r)$ \cite{besag1977_lfun,silverman1978, ripley1976}.

\begin{equation}
    \hat{L(r)} = r = \sqrt{\frac{\hat{E(r)}}{\lambda \pi}} = \frac{\sqrt{\nu(\Omega)}}{n} \sqrt{\sum_{i=1}^n \sum_{j \neq i}w_{ij} I(d_{ij} \leq r)}
    \label{eq:linearK}
\end{equation}

The linear transformation of Ripley's K-function, $\hat{L}(r)$, encodes information on the local interactions between points. In applied settings, researchers use the value of $r^{'}$ that maximizes $\hat{L}(r)$ to identify the radius of maximal aggregation, or the region that contains the most points per area, on average, around a test point \cite{kiskowski2009}. This provides a way to describe the clustering tendencies between points. 

The plug-in estimators introduced here do not encode information on local point interactions nor do they make implicit comparisons between observed and randomized patterns. Rather, these estimators search through the joint space $(\lambda,r)$ and identify the point $(\lambda(a_i), \hat{r})$ that maximizes a Palm likelihood function constrained to $k$\footnote{This does not include window adaptive perimeters}. Rather than identifying a radius of maximal aggregation, this constrained optimization selects the radius whose ball region "captures" k units, on average, across an infinite number of realizations. 

If $\lambda(a_i)$ is identified - $E[\lambda(a_i)] = \lambda$, then the plug-in estimators return an exact point estimate of $\hat{r}$. Across a finite sample of realizations, the estimators generate a sampling distribution for $\hat{r}$, which, under the right conditions, converges to a normal distribution centered at $r$. The Ripley's K-function, on the other hand, assigns a uniform distribution to $r$ on the positive Real line. Researchers sample locations from this uniform distribution to estimate $\hat{K(r)}$. When converted into the L-function, researchers can assess sampling fluctuations across different thresholds for $r$ (Besag 1977). By construction, these sampling fluctuations differ from the sampling distribution of $r$.

%% file: 3_Appendicies/Appendix_Scope_Analysis/C_Smallest_Bounding_Sphere_Problem.tex
\paragraph*{Smallest Bounding Sphere Problem}
Previous work on the smallest bounding sphere (SBS) is an optimization problem that shares superficial characteristics with the problem of dynamic spatial sampling \cite{yang2023,xu2003}. The SBS problem, first posed by James J. Sylvester in the 19th century, seeks to identify the smallest circular region that bounds $n$ points distributed in $\mathbb{R}^d$. \citeauthor[2003]{xu2003} state the optimization problem as follow: "Given a set of circles $C = \{c_1, ..., c_n\}$ on the Euclidean plane with centers $\{(a_1, b_1), ..., (a_n, b_n)\}$ and radii $\{r_1, ..., r_n\}$," what is "the circle of minimum radius that encloses all circles in $C$." Various algorithms have been proposed to handle this optimization problem \cite{fischer2004}. Within the set of scattered points $C$, there exists an implicit fixed point location whose circle radius satisfies this optimization problem. When solving the SBS problem, researchers use observed test points to identify an implicit fixed point location. The problem of DSS, on the other hand, establishes a fixed reference point, which can include a test point, and identifies the minimial sampling region that maximizes a Palm likelihood function constrained to $k$.

%% file: 3_Appendicies/Appendix_Proofs.tex
\section{Appendix 3: Proofs}
\label{appx:Proofs}

\subsection*{Proof A: Geofence Size and the Total Number of "Captured" Individuals}
\label{proofA}
\begin{proof}
    Here, we will prove that the number of captured agents $\hat{k}$ from the random process $\eta_X$ depends on the size of the geofence $D = B(a_i,r)$. We can frame agents intersecting with the geofence as a thinning process. Specifically, we define this thinning process as a series of Bernoulli trials defined across the random process $\eta$ \cite[p.158]{chiu2013}. Throughout this proof, we assume that points are uniformly distributed and that the probability that the point $x_i \in \eta_X$ lands inside the zone $D$ does not depend on the distribution of other points. 
    
    With this setup, for each point $x_i \in \eta_X$, we associate the binary indicator variable $\mathbbm{1}_{D}$, where:
    \begin{itemize}
        \item $P(a_i \in D) = \frac{\nu(B(a_i,r))}{\nu(\Omega)}$
        \item $P(a_i \notin D) = 1 -\frac{\nu(B(a_i,r))}{\nu(\Omega)}$
    \end{itemize}

    \noindent The characteristic function for this Bernoulli trial is given by:
    \begin{equation}
        \begin{aligned}
           E \bigg [e^{-\mathbbm{1}_{D}f(a_i)} \bigg ] &= \bigg (1-\frac{\nu(B(a_i,r))}{\nu(\Omega)}\bigg) + \frac{\nu(B(a_i,r))}{\nu(\Omega)}e^{-f(x)} \\
           &= 1 - \frac{\nu(B(a_i,r))}{\nu(\Omega)} \bigg (1 - e^{-f(x)}\bigg)
        \end{aligned} 
    \end{equation}
    
    \noindent Let $\eta_T$ represent the thinned process. The sum of the thinned process can be rewritten as a sum over the original process $\eta_X$:
    \begin{equation}
        \sum_{y_i\in\eta_T}f(y_i) = \sum_{a_i\in\eta}\mathbbm{1}_{D}f(a_i)
    \end{equation}

    \noindent We define the Laplace functional for the thinned process as \cite[p.284]{chiu2013}:
    \begin{equation}
        L(f)_{\eta_T} = E \bigg [ e^{-\sum_{a_i\in\eta}\mathbbm{1}_{D}f(a_i)}\bigg ]
    \end{equation}

    \noindent We re-define our Laplace functional using the product property of exponents:
    \begin{equation}
        L(f)_{\eta_T} = E \bigg [ \prod_{a_i\in\eta} e^{-\mathbbm{1}_{D}f(a_i)}\bigg ]
    \end{equation}

    \noindent The law of expectation allows us to re-define the Laplace functional. First, w take the expectation of the product of functionals conditioned on the locations in $\eta$. Second, we take the expected value across the entire process $\eta$. 
    \begin{equation}
        L(f)_{\eta_T} = E_{\eta} \bigg [ E_{D} \bigg [ \prod_{a_i\in\eta} e^{-\mathbbm{1}_{D}f(a_i)} \bigg | \eta\bigg ] \bigg ]
    \end{equation}
    
    \noindent Since the thinning decisions are independent, the inner expectation of the product becomes the product of expectations:
    \begin{equation}
        E_{D} \bigg [ \prod_{a_i\in\eta} e^{-\mathbbm{1}_{D}f(a_i)} \bigg ] =  \prod_{a_i\in\eta} E_{D} \bigg [e^{-\mathbbm{1}_{D}f(a_i)} \bigg ]
    \end{equation}

    \noindent  We can substitute the characteristic function for a single Bernoulli trial into the Laplace functional:
    \begin{equation}
        L(f)_{\eta_T} = E_{\eta} \bigg [  \prod_{a_i\in\eta} \bigg (1 - \frac{\nu(B(a_i,r))}{\nu(\Omega)} \bigg (1 - e^{-f(x)}\bigg)\bigg )\bigg ]
    \end{equation}

    \noindent  Recall the Probability Generating Functional (PGFL) of a PPP \cite[p.125]{chiu2013}, which states:
    \begin{equation}
        E \bigg [ \prod_{a_i\in\eta} h(a_i) \bigg ] = e^{-\int_{\mathbb{R}^d}(1-h(x))\Lambda(dx)}
    \end{equation}
    
    \noindent  We therefore will swap \(h(x)\) with the characteristic function: $1 - \frac{\nu(B(a_i,r))}{\nu(\Omega)} \bigg (1 - e^{-f(x)}\bigg)$.

    \noindent  Plugging in this value into the Laplace functional across all points, which gives us: 
    \begin{equation}
        L(f)_{\eta_T} = \text{exp} \Bigg (-\int_{\mathbb{R}^d}\bigg(1-\bigg [1-\frac{\nu(B(a_i,r))}{\nu(\Omega)} \bigg (1 - e^{-f(x)} \bigg) \bigg]\bigg)\Lambda(dx) \Bigg )
    \end{equation}

    \noindent Finally, simplifying the term inside the integral gives us:
    \begin{equation}
        L(f)_{\eta_T} = \text{exp} \Bigg (-\int_{\mathbb{R}^d}\bigg(1 - e^{-f(x)}\bigg)\frac{\nu(B(a_i,r))}{\nu(\Omega)}\Lambda(dx) \Bigg )
    \end{equation}

    \noindent Recall that the Laplace functional for a Poisson process is:
    \begin{equation}
                L(f)_{\eta_T} = \text{exp} \Bigg (-\int_{\mathbb{R}^d}\bigg(1 - e^{-f(x)}\bigg)\Lambda(dx) \Bigg )
    \end{equation}

    \noindent The proof shows that the thinned process is also a Poisson process. The number of captured individuals is a Poisson distributed random variable where the intensity of the process depends directly on the size of the catchment zone. If agents are uniformly distributed, the size of the zone is linearly related to the random summation output.
    \begin{equation}
        \Lambda_{\eta_T}(dx) = \frac{\nu(B(a_i,r))}{\nu(\Omega)}\Lambda(dx)
    \end{equation}
\end{proof}

\subsection*{Proof B: Privacy Constraint and the Total Number of "Captured" Individuals}
\label{proofB}
\begin{proof}
    Here, we will prove that for the focal adaptive plug-in estimator, the number of captured agents $\hat{k}$ from the random process $\eta_X$ depends only on the privacy constraint $k$. Let $\eta_X$ be a Poisson point process that is thinned by the zone $D = B(a_i,r)$. We can write its Laplace functional as:
    \begin{equation}
        L(f)_{\eta_T} = \text{exp} \Bigg (-\int_{\mathbb{R}^d}\bigg(1 - e^{-f(x)}\bigg)\frac{\nu(B(a_i,r))}{\nu(\Omega)}\lambda(x)dx \Bigg )
    \end{equation}

    \noindent Let the geofence $B(a_i,r)$ be an adaptive perimeter whose radius is defined by the focal adaptive estimator. We measure the size of the circular zone $\nu(B(a_i,r))$ as $\pi \hat{r}_{\text{Pois}}^2$:
    \begin{equation}
        L(f)_{\eta_T} = \text{exp} \Bigg (-\int_{\mathbb{R}^d}\bigg(1 - e^{-f(x)}\bigg)\frac{\pi \bigg (\sqrt{\frac{k}{\pi \lambda'_X (x)}} \bigg)^2}{\nu(\Omega)}\lambda(x)dx \Bigg )
    \end{equation}

    \noindent Simplifying the equation reduces it to: 
    \begin{equation}
        L(f)_{\eta_T} = \text{exp} \Bigg (-\int_{\mathbb{R}^d}\bigg(1 - e^{-f(x)}\bigg)\frac{k\pi}{\nu(\Omega)\pi \lambda'_X (x)}\lambda_X(x)dx \Bigg )
    \end{equation}

    \noindent Assuming no measurement error when estimating $\lambda'_X (x)$, then $\lambda'_X (x) = \lambda_X (x)$. Simplifying the equation further reduces it to: 
    \begin{equation}
        L(f)_{\eta_T} = \text{exp} \Bigg (-\int_{\mathbb{R}^d}\bigg(1 - e^{-f(x)}\bigg)\frac{k}{\nu(\Omega)}dx \Bigg )
    \end{equation}

    \noindent Holding fixed the size of the study window $\nu(\Omega)$, the intensity measure of the adaptive zone depends only on the parameter $k$. Multiplying the intensity measure by the size of the study window $\nu(\Omega)$ returns the total number of points across the study window, which equals $k$. 
    \begin{equation}
        \Lambda_{\eta_T}(dx) = \frac{k}{\nu(\Omega)}
    \end{equation}
\end{proof}

\subsection*{Proof C: Proof that the Sampling Distribution of $\hat{k}$ is  Normally Distributed}
\label{proofC}
\begin{proof}
    \noindent Let $\eta$ be a stationary Poisson point process thinned by the zone $D =B(a_i,r)$. It's intensity measure, therefore, is given by $\mu = \lambda \cdot \nu(B(a_i,r)$. Here, $\lambda$ is the intensity function of the stationary process, $\nu(B(a_i,r))$ is the size of the zone, and $\nu()$ is the Lebesgue integral. 

    \begin{equation}
        E[\eta(D)] = \lambda \cdot \nu(B(a_i,r))
    \end{equation}

    \noindent Since the process is Poisson distributed, the variance is given by:

    \begin{equation}
        Var[\eta(D)] = \lambda \cdot \nu(B(a_i,r))
    \end{equation}

    \noindent Let $Z_n$ be a standardized z-score for the intensity measure of the zone $D$:

    \begin{equation}
        Z_n = \frac{\eta(D) - E[\eta(D)]}{\sqrt{Var[\eta(D)]}} = \frac{\eta(D) - \mu}{\sqrt{\mu}}
    \end{equation}

    \noindent We will demonstrate how $Z_n$ converges to a standard normal distribution when $\mu \rightarrow \infty$ (or when  $\nu(B(a_i,r)) \rightarrow \infty$ or $\lambda \rightarrow \infty$)

    \noindent The moment generating function for a Poisson distribution is given by:

    \begin{equation}
        M_{\eta}(t) = e^{\mu(e^t-1)} = e^{\lambda \cdot \nu(B(a_i,r))\cdot(e^t-1)}
    \end{equation}

    \noindent The scaling properties of the moment generating function $M_{a\eta+b}(t) = e^{bt}M_{\eta}(at)$ allows us to generate a moment generating function for $Z_n$ \cite[p.62]{casella1990}. Let $Z_n = \frac{\eta(D) - \mu}{\sqrt{\mu}} = \frac{\eta(D)}{\mu} - \sqrt{\mu}$. We will define $a = \frac{1}{\sqrt{\mu}}$ and $b = -\sqrt{\mu}$. 

    \begin{equation}
        M_{Z_n}(t) = e^{-\sqrt{\mu}t}M_{\eta}(\frac{t}{\sqrt{\mu}})
    \end{equation}

    \noindent Substituting in the Poisson mgf gives us the following:
    \begin{equation}
        \begin{aligned}
            M_{Z_n}(t) &= e^{-\sqrt{\mu}t}e^{\mu\bigg(e^{\frac{t}{\sqrt{\mu}}}-1\bigg)}
        \end{aligned}
    \end{equation}

    \noindent Simplifying this Equation gives us the following:
    \begin{equation}
        \begin{aligned}
            M_{Z_n}(t) &= e^{\mu\bigg (e^{\frac{t}{\sqrt{\mu}}}-1\bigg) -\sqrt{\mu}t} 
        \end{aligned}
    \end{equation}

    \noindent We use the Taylor series expansion for $e^x$, where we substitute $x = \frac{t}{\sqrt{\mu}}$:

        
    $$e^{t/\sqrt{\mu}} = 1 + \frac{t}{\sqrt{\mu}} + \frac{t^2}{2\mu} + O\left(\mu^{-3/2}\right)$$

    \noindent Now, substitute this expansion back into our expression for the exponent of $M_{Z_n}(t)$:
    
    $$exp\bigg( \mu \left[ \left(1 + \frac{t}{\sqrt{\mu}} + \frac{t^2}{2\mu} + O(\mu^{-3/2})\right) - 1 \right] -\sqrt{\mu} t \bigg)$$
    
    \noindent  Simplify the term in the brackets (the $1$ and $-1$ cancel out):
    
    $$exp\bigg(\mu \left[ \frac{t}{\sqrt{\mu}} + \frac{t^2}{2\mu} + O(\mu^{-3/2}) \right] -\sqrt{\mu} t\bigg)$$
    
    \noindent Distribute $\mu$:
    
    $$exp\bigg( \left( \sqrt{\mu} t + \frac{t^2}{2} + O(\mu^{-1/2}) \right) -\sqrt{\mu} t \bigg)$$

    \noindent The $-\sqrt{\mu} t$ and $+\sqrt{\mu} t$ terms cancel out:
    
    $$exp\bigg( \frac{t^2}{2} + O\left(\frac{1}{\sqrt{\mu}}\right) \bigg )$$
    
    \noindent Now we take the limit as $\mu \to \infty$:
    
    $$\lim_{\mu \to \infty} M_{Z_n}(s) = \lim_{\mu \to \infty} \exp\left(\frac{t^2}{2} + O\left(\frac{1}{\sqrt{\mu}}\right)\right)$$
    
    $$\lim_{\mu \to \infty} M_{Z_n}(t) = e^{t^2/2}$$
    
    \noindent The function $e^{t^2/2}$ is the Moment Generating Function of the Standard Normal Distribution $\mathcal{N}(0, 1)$. By the Lévy's Continuity Theorem, if the MGF of a sequence of random variables converges to the MGF of a standard normal distribution, then the random variables converge in distribution to the standard normal distribution. Therefore, for large $\lambda \cdot \nu(B(a_i,r)$:
    
    $$N(t) \approx \mathcal{N}(\lambda \cdot \nu(B(a_i,r), \lambda \cdot \nu(B(a_i,r))$$
\end{proof}

%% file: 3_Appendicies/Appendix_Window.tex
\section{Appendix 4: Window Adaptive Plug-In Estimator}
\label{appx:window}
Let $\eta =\{x_i\}_{i=1}^m$ be a Binomial point process where $m$ points are uniformly distributed over the study window $\Omega$. Next, let $B \subseteq \Omega$ be our window adaptive geofence perimeter. Using this setup, let's begin with the simplest case: \textbf{a single Bernoulli trial where a randomly spaced point $x_i$ falls inside the geofence perimeter $B \subseteq \Omega$.} Equation \ref{eq:a1} defines the probability ($p$) that this point falls  inside the geofence perimeter. One can calculate this probability by dividing the size of the geofence region $\nu(B)$ by the size of the jurisdictional study window $\nu(\Omega)$, where $\nu$ is a Lebesgue measure. Increasing the size of our geofence increases the probability that one individual falls within its boundaries. We can define the probability ($q$) that the individual \textit{does not} fall inside the perimeter by dividing the size of the perimeter's complement with the total size of the window. 
\begin{equation}
    \begin{split}
        p &= P(x_i \in B) = \frac{\nu(B)}{\nu(\Omega)} \\
        \\
        q &= P(x_i \notin B) = \frac{\nu(\Omega \text{\textbackslash} B)}{\nu(\Omega)} = 1 - \frac{\nu(B)}{\nu(\Omega)}
    \end{split}
    \label{eq:a1}
\end{equation}
    
We can extend this probability into the conditional realm where we condition on the focal surveillance site $a_i \in \Omega$. Equation \ref{eq:a2} summarizes the probability that a point falls, or does not fall, into the perimeter $B(a_i,r)$ centered at the fixed point location $a_i$. Similar to Equation \ref{eq:a1}, we calculate this probability ($p$) by dividing the size  of $\nu(B(a_i,r))$ by the total area of the study window $\nu(\Omega)$. We calculate $q$ by taking its complement. 

\begin{equation}
    \begin{split}
        p &= P(x_i \in B(a_i,r)) = \frac{\nu(B(a_i,r))}{\nu(\Omega)} \\
        \\
        q &= P(x_i \notin B(a_i,r)) = 1 - \frac{\nu(B(a_i,r))}{\nu(\Omega)}
    \end{split}
    \label{eq:a2}
\end{equation}

With this setup, we can use the Binomial distribution to estimate the probability that $k$ individuals will intersect with $B(a_i,r)$. Let $N$ be a random counting measure and $N(B(a_i,r))$ be the number of points that intersect with the ball region $B(a_i,r)$, then Equation \ref{eq:a4} gives the probability that exactly $k$ points intersect with $B(a_i,r)$. In addition to our selection for $k$, this probability depends heavily on the size of the study window $\nu(\Omega)$ and the total population inside the window $n$.

\begin{equation}
    P(N(B(a_i,r)) = k) = {n \choose k} \biggr[\frac{\nu(B(a_i,r))}{\nu(\Omega)}\biggr]^k\biggr[1 - \frac{\nu(B(a_i,r))}{\nu(\Omega)}\biggr]^{n-k}
    \label{eq:a4}
\end{equation}

Using the information from Equation \ref{eq:a4}, we can use the maximum likelihood estimator for the binomial distribution to identify an optimal radius $\hat{r}$ and sampling region around $a_i$ that returns $k$ points, on average. Equation \ref{eq:a5} describes the log-likelihood for the Binomial distribution in terms of the spatial quantities. Let $p = \frac{\nu(B(a_i,r))}{\nu(\Omega)}$, then taking the derivative of Equation \ref{eq:a5} and solving with respect to $p$ generates the maximum likelihood estimator $\hat{p}$. Setting $\hat{p} = \frac{\nu(B)}{\nu(\Omega)}$ produces the equality relationship in Equation \ref{eq:a6}. It assumes the proportion of area coverage between the geofence perimeter $\nu(B(a_i,r))$ and study window $\nu(\Omega)$ equals the proportion of captured individuals. This holds when points are independently and uniformly distributed across the study window. 

\begin{equation}
    \mathcal{LL}(p) = log {n \choose k} + k \;log \bigg (\frac{\nu(B(a_i,r))}{\nu(\Omega)} \bigg ) + (n-k)\;log \bigg (1- \frac{\nu(B(a_i,r))}{\nu(\Omega)} \bigg )
    \label{eq:a5}
\end{equation} 

\begin{equation}
    \frac{d\mathcal{LL}(p)}{dp} = \frac{k}{p} - \frac{n-k}{1-p} = 0
\end{equation}

\begin{equation}
   \hat{p} = \frac{\nu(B)}{\nu(\Omega)} = \frac{k}{n} 
   \label{eq:a6}
\end{equation}

We can re-arrange the left side of Equation \ref{eq:a6} to generate a function where the size of the geofence perimeter $\nu(B(a_i,\hat{r}))$ equals the size of the study window $\nu(\Omega)$ multiplied by the proportion of captured individuals $\frac{k}{n}$. To generate the optimal radius estimator, we have to define the functional form of $\nu(B)$. We will focus on a circular geofence perimeter where we substitute $\nu(B)$ with $\pi r^2$. Solving for $r$ produces the window adaptive plug-in estimator (see Equation \ref{eq:a7}).  

\begin{equation}
    \frac{\nu(B(a_i,r))}{\nu(\Omega)} = \frac{k}{n} \leadsto \nu(B(a_i,\hat{r})) = \nu(\Omega)\frac{k}{n}
\end{equation}

\begin{equation}
    \pi \hat{r}^2 = \nu(\Omega)\frac{k}{n} \leadsto r = \sqrt{\nu(\Omega)\frac{k}{n\pi}}
    \label{eq:a7}
\end{equation}

\begin{figure}
    \centering
    \includegraphics[width=0.95\linewidth]{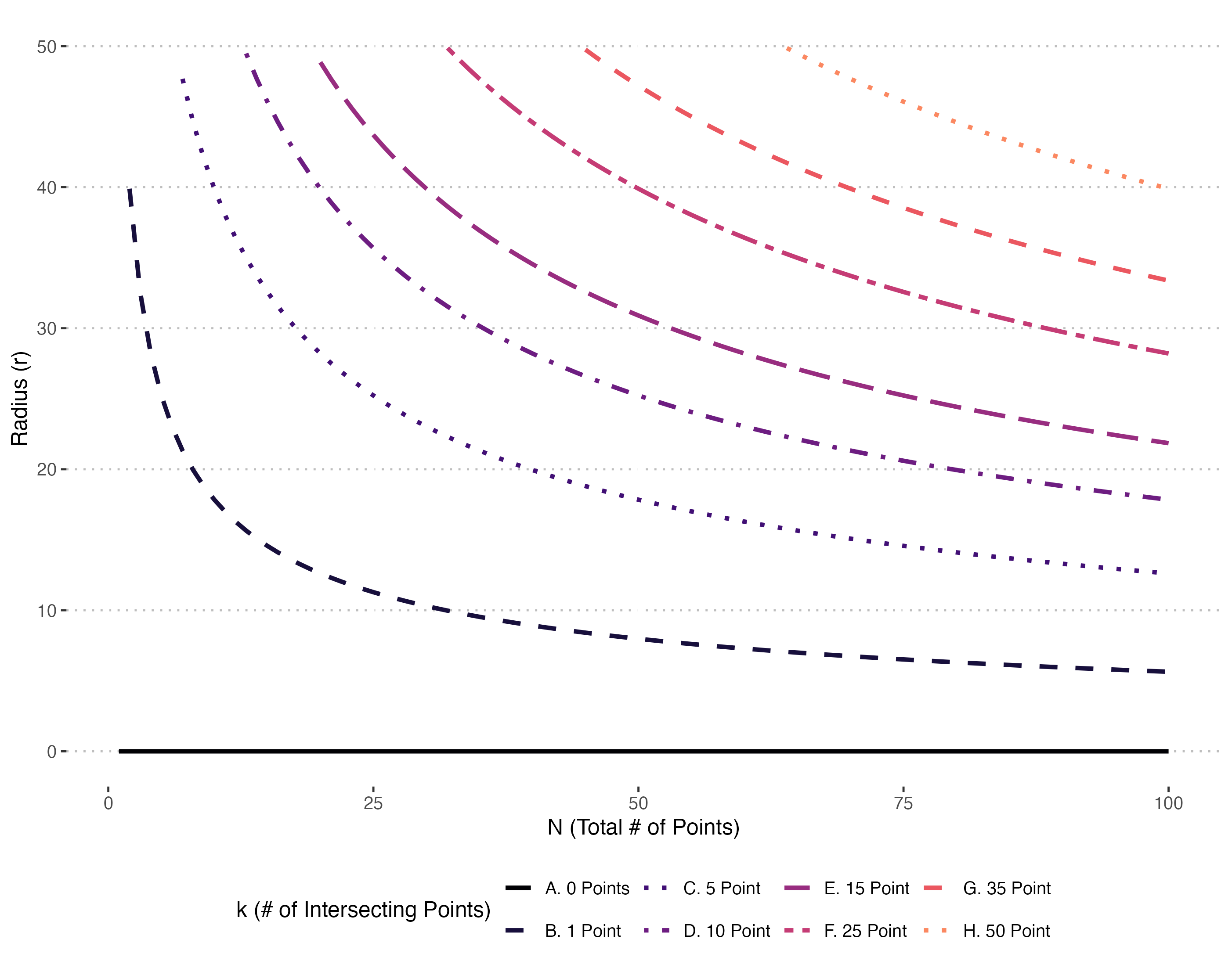}
    \caption{Radius ($r$) and Number of Commuters ($n$) across different Privacy Constraints ($k$)} 
    \label{fig:optimalWindow}
\end{figure}

Figure \ref{fig:optimalWindow} illustrates the non-linear relationship between the total number of points  $n$ and the optimal ball radius $\hat{r}$ across different total number of intersecting points $k$. We fix the study window to $[0,100]^2$. Each line represents a specific boundary condition where the estimator $\hat{r}$ satisfies the privacy constraint $k$. Shifting away from the boundary biases the expected number of "captured" individuals. 

\begin{figure}
    \centering
    \includegraphics[width=0.95\linewidth]{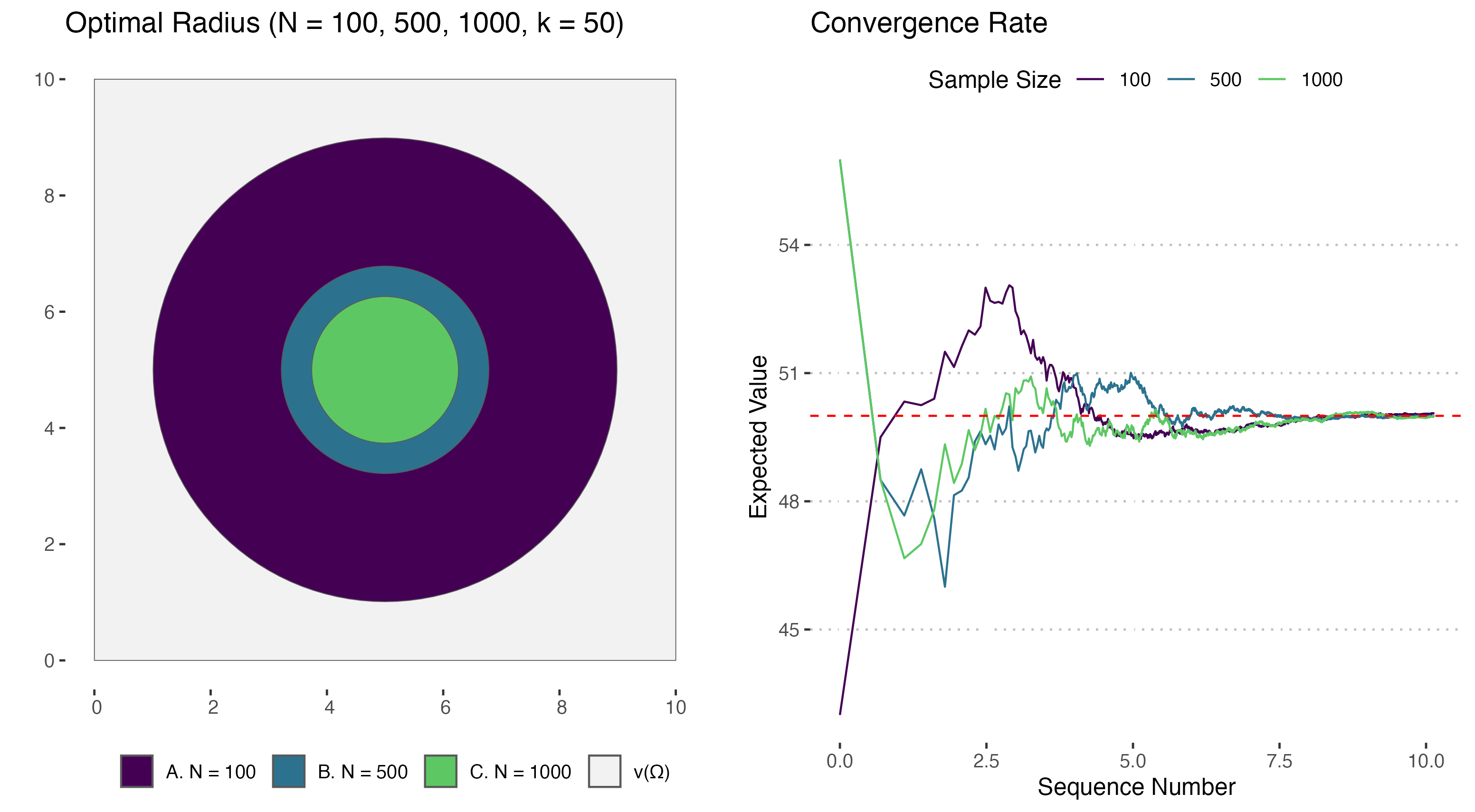}
    \caption{Window Adaptive Sampling Regions and Convergence Rates for $k = 50$}
    \label{fig:convergenceWindow}
\end{figure}

Figure \ref{fig:convergenceWindow} visualizes this non-linear relationship between sampling ball region and number of commuters for a study window with size $[0,100]^2$. Here, we set the privacy constraint to 50 commuters; however, we vary the total number of commuters, where $n = \{100, 500, 1000\}$. The left graph illustrates the sampling regions while the right graph visualizes how the expected number of "captured" individuals converges to the privacy constraint ($k = 50$). 

%% file: 3_Appendicies/Appendix_Focal.tex
\section{Appendix 5: Focal Adaptive Plug-In Estimator}
\label{appx:focal}
Let \(\eta =\{x_i\}_{i=1}^{\infty}\) be a stationary Poisson point process with a density of $\lambda$. Integrating $\lambda$ across all subsets of the study window $\omega \subseteq \Omega$ returns the total number of points inside the study window $\Lambda(\Omega)$. This population total is assumed to be drawn from a Poisson distribution whose expected value is $\Lambda(\Omega)$. Let $B(a_i, r)$ be a local neighborhood inside the study window $B \subseteq \Omega$ whose surveillance site equals $a_i$. Finally, let $\eta(\{B(a_i, r)\})$ be the point locations that belong to the process $\eta$ inside the area $B \subseteq \Omega$. 

\begin{equation}
    P(\eta(\{B(a_i, r)\}) > 0) = 1 - P(\eta(\{B(a_i, r)\}) = 0) = 1 - e^{-\lambda \nu(B(a_i, r))}
    \label{eq:b1}
\end{equation}

Using this setup, Equation \ref{eq:b1} describes the probability that at least one point falls within a given area $B(a_i, r) \subseteq \Omega$. We can use Equation \ref{eq:b2} to calculate the probability that $k$ points fall within the region $\nu(B(a_i, r))$. 

\begin{equation}
    P(N(B(a_i, r)) = k) = \frac{[\lambda \nu(B(a_i, r))]^k}{k!}e^{-\lambda \nu(B(a_i, r))} = \frac{[\lambda \pi r^2]^k}{k!}e^{-\lambda \pi r^2}
    \label{eq:b2}
\end{equation}

Similar to the window adaptive case, we can use the Poisson Distribution's maximum likelihood estimator to identify an optimal radius $\hat{r}$ and sampling region around $a_i$ that returns an unbiased and consistent estimate of $k$ points (see Equation \ref{eq:b3}). Solving for $\lambda$ and $r$ return the following estimators for the optimal intensity $\hat{\lambda}$ and the optimal radius $\hat{r}$, where $k$ represents the sample mean $\frac{1}{n} \sum_{i=1}^n x_i$. 

\begin{equation}
    \mathcal{LL}(\lambda,r) = -n\lambda \pi r^2 - \sum_{i=1}^{n}ln(x_i!) + ln(\lambda \pi r^2) \sum_{i=1}^n x_i
    \label{eq:b3}
\end{equation}

These two estimators identify the local maxima in the joint space ($\lambda$, $r$) that maximizes the conditional Palm distribution (see Equation \ref{eq:b4}). In this context, the optimal intensity normalizes the sample mean by the size of the sampling region. The optimal radius, on the other hand, normalizes the sample mean by the spherical intensity $\pi \lambda$ and takes its square root. The vector $[\hat{\lambda,\hat{r}]}$ is asymptotically normal. 

\begin{equation}
    \hat{\lambda} = \frac{1}{r^2}\frac{\sum_{i=1}^n x_i}{n \pi} = \frac{k}{\pi r^2} \quad \quad \quad \quad \quad \quad \quad \quad  \hat{r} = \sqrt{\frac{\sum_{i=1}^n x_i}{n \lambda \pi}} = \sqrt{\frac{k}{\pi \lambda}} 
    \label{eq:b4}
\end{equation}

\paragraph{Partial-Derivative for $\lambda$:}
\begin{equation} 
    \begin{split}
        \frac{\partial}{\partial\lambda}\mathcal{LL}(\lambda,r) & =  \frac{\partial}{\partial\lambda} \Biggr[ -n\lambda \pi r^2 - \sum_{i=1}^{n}ln(x_i!) + ln(\lambda \pi r^2) \sum_{i=1}^n x_i \Biggr] \\
                                                            & = \frac{\sum_{i=1}^n x_i}{\lambda} - \pi n r^2 
    \end{split}
\end{equation}

\begin{equation} 
    \begin{split}
        \frac{\sum_{i=1}^n x_i}{\lambda} - \pi n r^2  & = 0 \\
                                                    \frac{\sum_{i=1}^n x_i}{\lambda} & = \pi n r^2 \\
                                                    \sum_{i=1}^n x_i & = \lambda \pi n r^2 \\
                                                    \frac{\sum_{i=1}^n x_i}{\pi n r^2} & = \lambda  \\
    \end{split}
\end{equation}

\paragraph{Partial-Derivative for $r$:}
\begin{equation} 
    \begin{split}
        \frac{\partial}{\partial r}\mathcal{LL}(\lambda,r) & =  \frac{\partial}{\partial r} \Biggr[ -n\lambda \pi r^2 - \sum_{i=1}^{n}ln(x_i!) + ln(\lambda \pi r^2) \sum_{i=1}^n x_i \Biggr] \\
                                                            & = -2n\lambda \pi r + \frac{2 \lambda \pi r}{\lambda \pi r^2}\sum_{i=1}^n x_i \\
                                                            & = -2n\lambda \pi r + \frac{2}{r}\sum_{i=1}^n x_i
    \end{split}
\end{equation}

\begin{equation} 
    \begin{split}
        -2n\lambda \pi r + \frac{2}{r}\sum_{i=1}^n x_i & = 0 \\
                                                     \frac{2}{r}\sum_{i=1}^n x_i    & = 2n\lambda \pi r \\
                                                     \frac{1}{r}\sum_{i=1}^n x_i    & = n\lambda \pi r \\
                                                     \sqrt{\frac{\sum_{i=1}^n x_i}{n\lambda \pi}}    & = r
    \end{split}
\end{equation}

\paragraph{Score Vector}
\begin{equation}
    \nabla_{\lambda,r}ln(f_X(X_j;\lambda,r)) = \begin{bmatrix}
                                                        \frac{X_j}{\lambda} - \pi n r^2 \\
                                                        -2n\lambda \pi r + \frac{2}{r}X_j
                                                \end{bmatrix}
\end{equation}

\paragraph{Hessian Matrix}
\begin{equation}
    \nabla^2_{\lambda,r}ln(f_X(X_j;\lambda,r)) = \begin{bmatrix}
                                                        \frac{\partial^2}{\partial \lambda^2} ln(f_X(X_j;\lambda,r)) & \frac{\partial}{\partial \lambda} \frac{\partial}{\partial r} ln(f_X(X_j;\lambda,r)) \\
                                                        \frac{\partial}{\partial r} \frac{\partial}{\partial \lambda} ln(f_X(X_j;\lambda,r) & \frac{\partial^2}{\partial r^2} ln(f_X(X_j;\lambda,r))
                                                \end{bmatrix}
\end{equation}

\begin{equation} 
    \begin{split}
        \frac{\partial^2}{\partial \lambda^2} ln(f_X(X_j;\lambda,r))  & = \frac{\partial^2}{\partial \lambda^2} \Biggr[ -n\lambda \pi r^2 - \sum_{i=1}^{n}ln(x_i!) + ln(\lambda \pi r^2) \sum_{i=1}^n x_i \Biggr]  \\
        & = -\frac{1}{\lambda^2}X_j
    \end{split}
\end{equation}

\begin{equation} 
    \begin{split}
        \frac{\partial^2}{\partial r^2} ln(f_X(X_j;\lambda,r))  & = \frac{\partial^2}{\partial r^2} \Biggr[ -n\lambda \pi r^2 - \sum_{i=1}^{n}ln(x_i!) + ln(\lambda \pi r^2) \sum_{i=1}^n x_i \Biggr]  \\
        & = -2n \lambda \pi - \frac{2}{r^2}X_j
    \end{split}
\end{equation}

\begin{equation} 
    \begin{split}
        \frac{\partial}{\partial \lambda} \frac{\partial}{\partial r} ln(f_X(X_j;\lambda,r))  & = \frac{\partial}{\partial \lambda} \frac{\partial}{\partial r} \Biggr[ -n\lambda \pi r^2 - \sum_{i=1}^{n}ln(x_i!) + ln(\lambda \pi r^2) \sum_{i=1}^n x_i \Biggr]  \\
        & = - 2\pi n r 
    \end{split}
\end{equation}

\begin{equation}
    \nabla^2_{\lambda,r}ln(f_X(X_j;\lambda,r)) = \begin{bmatrix}
                                                        -\frac{1}{\lambda^2}X_j & - 2\pi n r  \\
                                                        - 2\pi n r  & -2n \lambda \pi - \frac{2}{r^2}X_j
                                                \end{bmatrix}
\end{equation}

\noindent \textbf{Information Equality}
\begin{equation}
    \begin{split}
        Var[\nabla^2_{\lambda,r}ln(f_X(X_j;\lambda,r))] & = \begin{bmatrix}
                                                        -\frac{1}{\lambda^2}X_j & - 2\pi n r  \\
                                                        - 2\pi n r  & -2n \lambda \pi - \frac{2}{r^2}X_j
                                                        \end{bmatrix} \\
                                                        & = -E[\nabla^2_{\lambda,r}ln(f_X(X_j;\lambda,r))] \\
                                                        & = \begin{bmatrix}
                                                        \frac{1}{\lambda^2}\lambda &  2\pi n r  \\
                                                         2\pi n r  & 2n \lambda \pi + \frac{2}{r^2}\lambda
                                                        \end{bmatrix} \quad \text{because} \quad E[X_j] = \lambda \\
                                                        & = \begin{bmatrix}
                                                        \frac{1}{\lambda} &  2\pi n r  \\
                                                         2\pi n r  & 2n \lambda \pi + \frac{2}{r^2}\lambda
                                                        \end{bmatrix}
    \end{split}
\end{equation}

\noindent \textbf{Asymptotic Covariance Matrix}
\begin{equation}
    \frac{1}{n}[Var[\nabla^2_{\lambda,r}ln(f_X(X_j;\lambda,r))]]^{-1} = \frac{1}{n}\begin{bmatrix}
                                                        \lambda & \frac{1}{ 2\pi n r}  \\
                                                        \frac{1}{ 2\pi n r}  & \frac{1}{2n \lambda \pi} + \frac{r^2}{2\lambda}
                                                        \end{bmatrix}
\end{equation}

\begin{figure}
    \centering
    \includegraphics[width=0.95\linewidth]{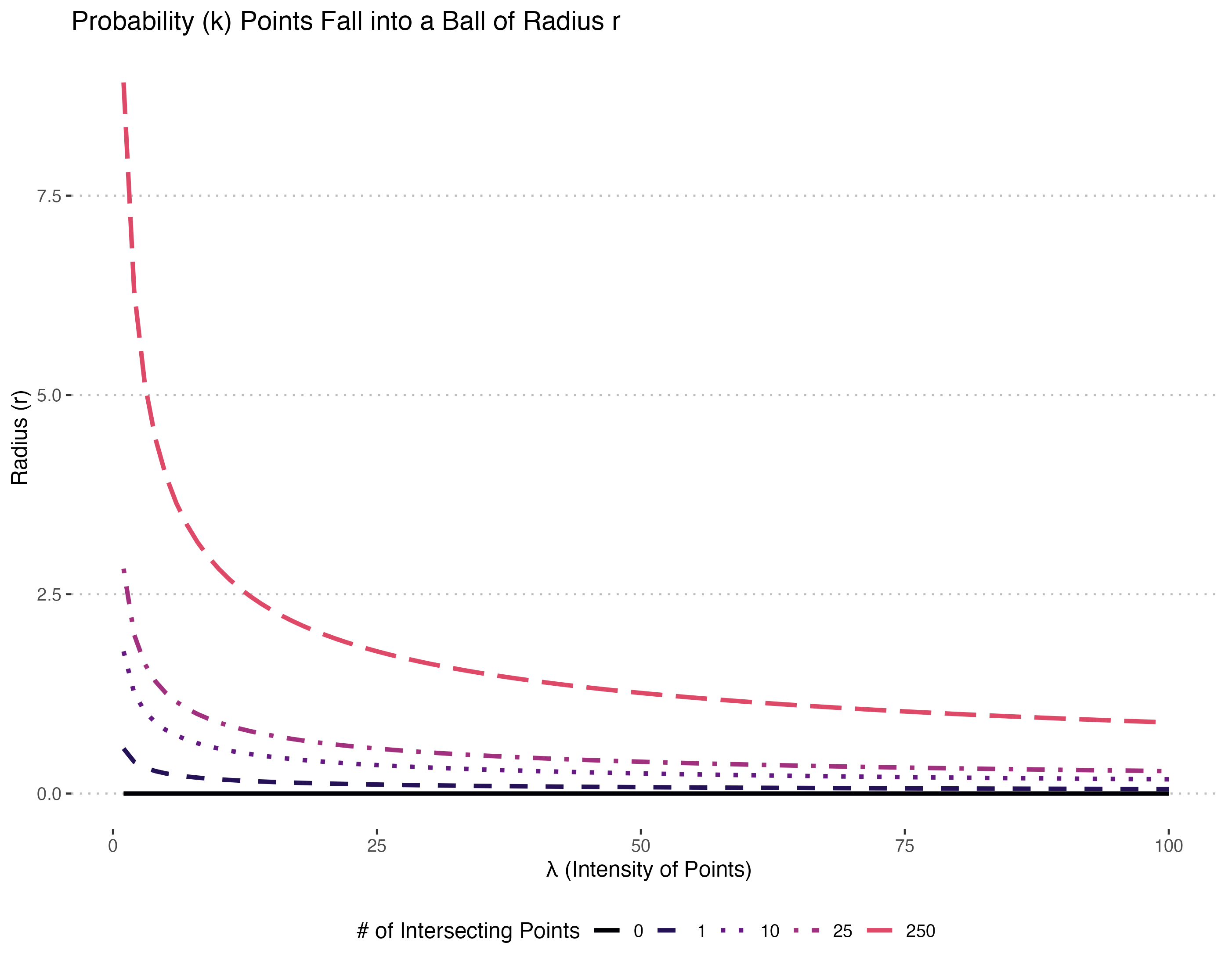}
    \caption{Finding the Optimal Ball Radius (Stationary Point Process)}
    \label{fig:optimalFocal}
\end{figure}

\begin{figure}
    \centering
    \includegraphics[width=0.95\linewidth]{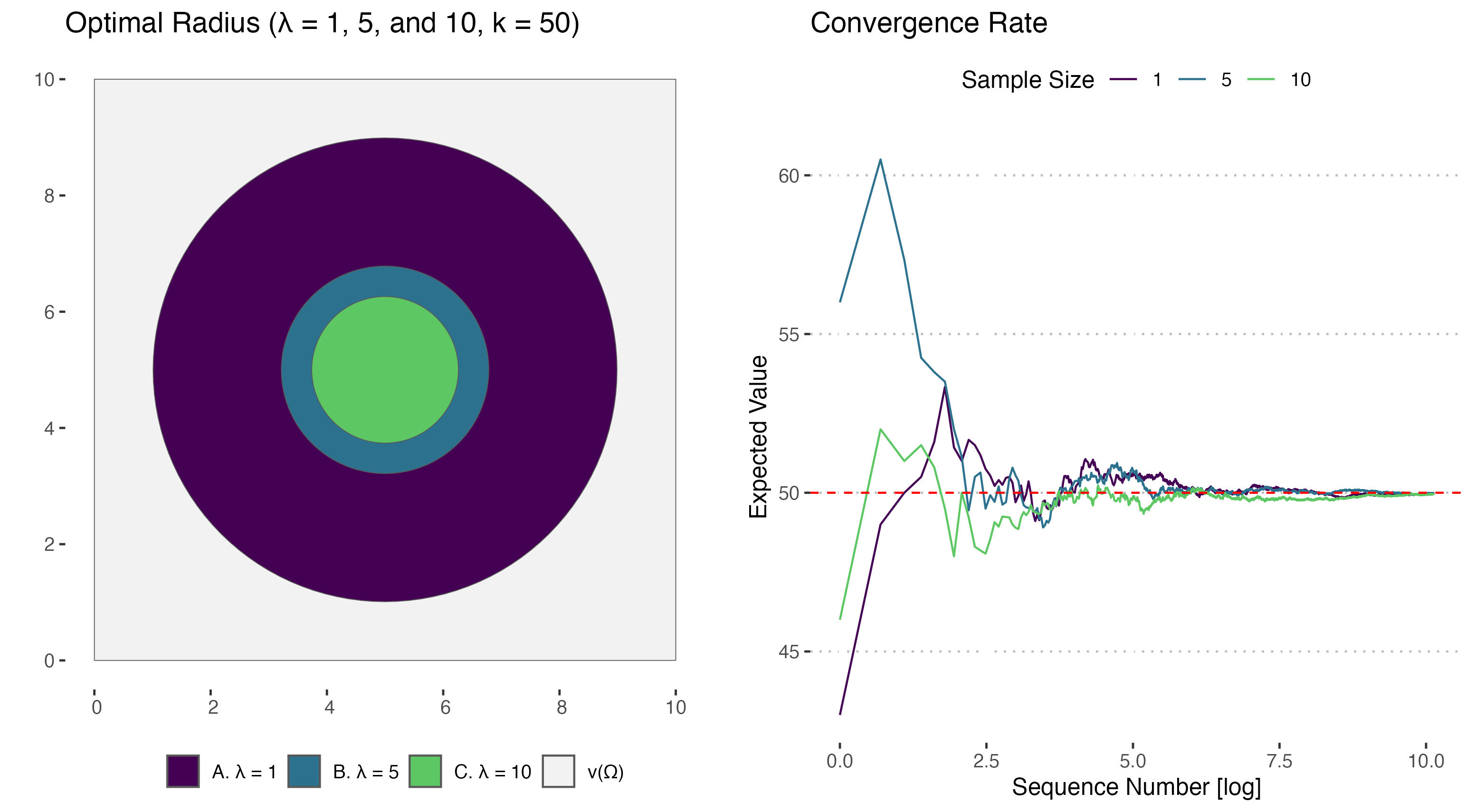}
    \caption{Convergence Rate for Optimal Radius: Stationary Point Process (k = 50)}
    \label{fig:convergenceFocal}
\end{figure}

In Figure \ref{fig:convergenceFocal}, we visualize the convergence rate for three sampling regions constrained to $k=50$. We assign different local densities - $\lambda(a_i) = \{1, 5, 10\}$ - to the fixed reference location $(5,5)$ inside the plane $[0,10]^2$. The expected number of commuters in the study window, therefore, equals $E[\lambda([0,10]^2)] = \{100,500,1000\}$ and are assumed to be Poisson distributed. The left graph illustrates the sampling regions while the right graph visualizes how the expected number of "captured" individuals converges to the privacy constraint. 

%% file: 3_Appendicies/Appendix_Lambda.tex
\section{Appendix 6: Lambda Adaptive Plug-In Estimator}
\label{appx:lambda}
The focal adaptive estimator assumes points are uniformly distributed across the study window, including inside the geofence perimeter.  We can use inhomogeneous Poisson point processes to relax this assumption and account for this non-stationarity. These processes assume points are independent but unevenly distributed across the study window.

We begin by defining the intensity for a inhomogeneous process (see Equation \ref{eq:c1}). Let \( \lambda(x) \) be the intensity function of a nonhomogeneous Poisson process across the geofence \(B = B(a_i,r) \subseteq \Omega \). Using this setup, we can define the log-likelihood function $\mathcal{LL}(\lambda(B), r)$ (see Equation \ref{eq:c2}). This function sums the intensity of each point that intersects with the ball region $B(a_i,r)$ and subtracts the expected number of points across the ball region. We add the multiplier term $\alpha$ to allow for multiple points to intersect with the ball region. 

\begin{equation}
    \Lambda(B) = \alpha\cdot\int_{B}\lambda(x)dx
    \label{eq:c1}
\end{equation}

\begin{equation}
    \mathcal{LL}(\lambda(B), r) = \alpha\cdot\sum_{i=1}^n \log \bigg (\lambda(x_i)\nu(B) \bigg ) - \int_S \lambda(x)\nu(B) \, dx.
    \label{eq:c2}
\end{equation}

 Since the integral term $\int_S \lambda(x)\nu(B) \, dx$ is analytically intractable, we use a quadrature scheme to discretize the integral term into a weighted summation term (Equation \ref{eq:c3}) (Berman and Turner, 1992). Here, $S$, $X$ and $S-X$ are the design points, the observed data points and the dummy points respectively. The weights $w_k$ are determined by some quadrature rule with a set of quadrature points $\{s_1, s_2, ..., s_n \}$. The observed points are a subset of the quadrature points, where  $\{x_1, x_2, ..., x_n \} \subset \{s_1, s_2, ..., s_n \}$\footnote{Mis-specification in the pseudo-likelihood estimate arises from the resolution of the discrete field. Increasing the resolution decreases the estimation bias as the manifolds surrounding each quadrature point converges towards zero.}. We can solve for the optimal radius by taking the partial derivative of Equation \ref{eq:c3}, which produces the estimator $\hat{r}_{\lambda}.$ According to Equation \ref{eq:c3}, the optimal radius $\hat{r}_{\lambda}$ depends on the sum of densities across the design points $S$. As the density across the design points increases, the optimal radius shrinks towards zero.

\begin{equation}
    \begin{split}
        \mathcal{LL}(\lambda) &= \alpha \cdot \sum_{i=1}^n \log \bigg ( \lambda(x_i)\nu(B) \bigg ) - \sum_{k = 1}^{|S|} w_k \, \lambda(s_k)\nu(B) \\
        &=  \alpha \cdot \sum_{i=1}^n \log \bigg (\lambda(x_i)\pi r^2 \bigg ) - \sum_{k = 1}^{|S|} w_k \, \lambda(s_k)\pi r^2
    \end{split}
\end{equation}

\begin{equation}
    \begin{split}
        \frac{\partial}{\partial r}\mathcal{LL}(\lambda) & = \frac{\partial}{\partial r}\biggr [\alpha \cdot\sum_{i=1}^n \log \big (\lambda(x_i)\pi r^2 \big ) - \sum_{k = 1}^{|S|} w_k \, \lambda(s_k)\pi r^2 \biggr] \\
        &= \alpha \cdot\frac{2\sum_{i=1}^n  \lambda(x_i)\pi r }{\sum_{i=1}^n  \lambda(x_i)\pi r^2} - 2\sum_{k = 1}^{|S|} w_k \, \lambda(s_k)\pi r
    \end{split}
\end{equation}

\begin{equation}
    \begin{split}
        \alpha \cdot\frac{2\sum_{i=1}^n  \lambda(x_i)\pi r}{\sum_{i=1}^n  \lambda(x_i)\pi r^2} - 2\sum_{k = 1}^{|S|} w_k \, \lambda(s_k)\pi r &= 0 \\
        \alpha \cdot\frac{2}{r}&=  2\sum_{k = 1}^{|S|} w_k \, \lambda(s_k)\pi r \\
        \alpha  &=  \sum_{k = 1}^{|S|} w_k \, \lambda(s_k) r^2 \\
        \sqrt{\frac{\alpha }{\sum_{k = 1}^{|S|} w_k \, \lambda(s_k) \pi}} &= \hat{r}_{\lambda}
        \label{eq:c3}
    \end{split}
\end{equation}

\begin{figure}
    \centering
    \includegraphics[width=0.95\linewidth]{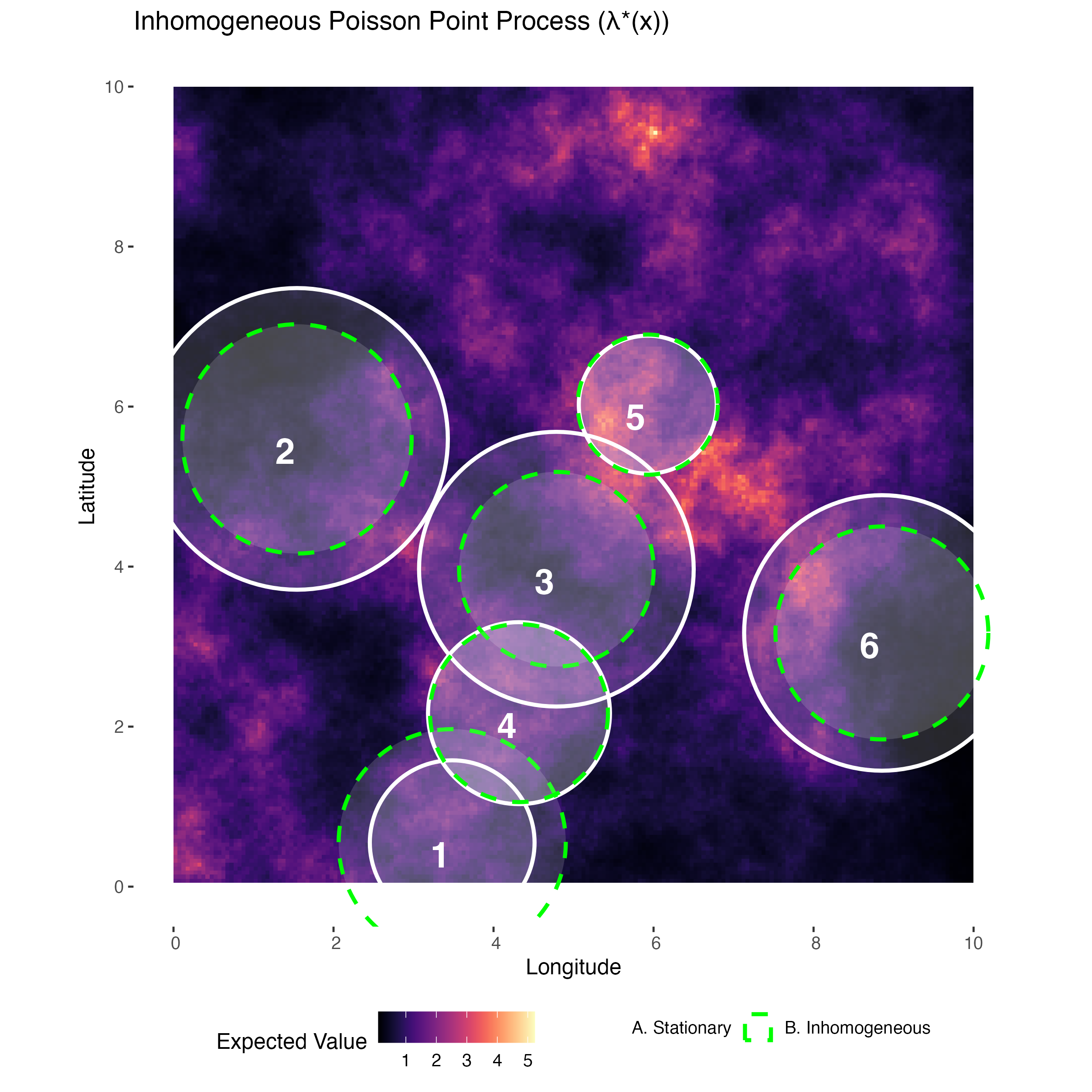}
    \caption{Finding the Optimal Ball Radius (Stationary Point Process)}
    \label{fig:optimalLambda}
\end{figure}

In Figure \ref{fig:optimalLambda}, we demonstrate how $\hat{r}_{\text{Pois}}$ (solid) and $\hat{r}_{\text{Inhom}}$ (dashed) differ when the local point densities inside the convex hull of $B(a_i, \hat{r}_{\text{Pois}})$ are unevenly distributed. First, we generate a random field with a spatially varying intensity function in $[0,10]^2$. Here, lightly shaded regions denote high density regions while darkly shaded regions denote low density regions. Next, we randomly select six focal surveillance points, measure their associated intensities, and use $\hat{r}_{\text{Pois}}$ to estimate circular geofence perimeters that return five ($k=5$) intersecting points, on average. We estimate $\hat{r}_{\text{Inhom}}$ and construct new sampling regions that adjust for the non-stationarity inside each geofence perimeters' convex hull.

 \begin{equation}
    \underset{r \in S}{\text{argmin}}\bigg [\bigg (\Lambda(B(a_i, \hat{r}_{\text{Pois}})) - k) \bigg )^2\bigg ]
\end{equation}

Since $\hat{r}_{\text{Inhom}}$ is weakly monotonic, the value of $\hat{r}_{\text{Inhom}}$ is not directly invertible with the local density at the focal surveillance $\lambda(a_i)$. In practice, three steps are required for identifying this value. We construct these perimeters in several stages. First, we estimate the local density at their focal surveillance site $\lambda(a_i)$. Second, we construct a focal adaptive perimeter subject to the privacy constraint $k$. Third, we estimate the number of "captured" individuals $\hat{k} = \Lambda(B(a_i, \hat{r}_{\text{Pois}}))$. Fourth, we sample several different radii $\epsilon$ from the interval $(-r, 2r]$, add $\epsilon$ to $\hat{r}_{\text{Pois}}$ (buffering), and then estimate the total number of "captured" individuals for each newly buffered perimeter. Finally, we adjust the perimeter size by minimizing an objective function that measures deviance between the estimate $\hat{k}$ and the original privacy constraint $k$.

%% file: 3_Appendicies/Appendix_Central.tex
\section{Appendix 7: Central Angle Adaptive Plug-In Estimator}
\label{appx:central}
We derive the central angle adaptive estimator by substituting $\nu(B(a_i,r))$ in the focal adaptive estimator with the central angle formula $A$ (see Equation \ref{eq:d1}). Equation \ref{eq:d2} describes the modified log-likelihood function.  We derive the partial derivatives for each of the three parameters $(\lambda,r,\theta)$.

\begin{equation}
    A = \frac{1}{2}r^2\theta
    \label{eq:d1}
\end{equation}

\begin{equation}
    \mathcal{LL}(\lambda,r,\theta) = \frac{-n\lambda r^2\theta}{2}  - \sum_{i=1}^{n}ln(x_i!) + ln(\frac{\lambda r^2\theta}{2}) \sum_{i=1}^n x_i
    \label{eq:d2}
\end{equation}

\paragraph{Partial-Derivative for $\lambda$:}
\begin{equation} 
    \begin{split}
        \frac{\partial}{\partial\lambda}\mathcal{LL}(\lambda,r,\theta) & =  \frac{\partial}{\partial\lambda} \Biggr[ \frac{-n\lambda r^2\theta}{2}  - \sum_{i=1}^{n}ln(x_i!) + ln(\frac{\lambda r^2\theta}{2}) \sum_{i=1}^n x_i \Biggr] \\
        & = \frac{-nr^2\theta}{2} + \frac{\sum_{i=1}^n x_i}{\lambda}
    \end{split}
\end{equation}

\begin{equation} 
    \begin{split}
        \frac{-nr^2\theta}{2} + \frac{\sum_{i=1}^n x_i}{\lambda}  & = 0 \\
        \frac{\sum_{ i=1}^n x_i}{\lambda} & = \frac{nr^2\theta}{2} \\
        \sum_{i=1}^n x_i & = \frac{n\lambda r^2\theta}{2} \\
        2\cdot\frac{\sum_{i=1}^n x_i}{n r^2\theta} & = \lambda  \\
    \end{split}
\end{equation}

\paragraph{Partial-Derivative for $r$:}
\begin{equation} 
    \begin{split}
        \frac{\partial}{\partial r}\mathcal{LL(\lambda,r,\theta)} & =  \frac{\partial}{\partial r} \Biggr[ \frac{-n\lambda r^2\theta}{2}  - \sum_{i=1}^{n}ln(x_i!) + ln(\frac{\lambda r^2\theta}{2}) \sum_{i=1}^n x_i \Biggr]  \\
        & = -\frac{2n\lambda  r\theta}{2} + \frac{4 \lambda  r\theta}{2\lambda  r^2\theta}\sum_{i=1}^n x_i \\
        & = -n\lambda  r\theta + \frac{2}{r}\sum_{i=1}^n x_i
    \end{split}
\end{equation}

\begin{equation} 
    \begin{split}
        -n\lambda  r\theta + \frac{2}{r}\sum_{i=1}^n x_i & = 0 \\
        \frac{2}{r}\sum_{i=1}^n x_i    & = n\lambda  r\theta \\
        \sqrt{2\cdot\frac{\sum_{i=1}^n x_i}{n\lambda \theta}}    & = r
    \end{split}
\end{equation}

\paragraph{Partial-Derivative for $\theta$:}
\begin{equation} 
    \begin{split}
        \frac{\partial}{\partial r}\mathcal{LL(\lambda,r,\theta)} & =  \frac{\partial}{\partial \
        \theta} \Biggr[ \frac{-n\lambda r^2\theta}{2}  - \sum_{i=1}^{n}ln(x_i!) + ln(\frac{\lambda r^2\theta}{2}) \sum_{i=1}^n x_i \Biggr]  \\
        & = -\frac{n\lambda  r^2}{2} + \frac{2\lambda r^2}{2\lambda  r^2\theta}\sum_{i=1}^n x_i \\
        & = -\frac{n\lambda  r^2}{2} + \frac{1}{\theta}\sum_{i=1}^n x_i
    \end{split}
\end{equation}

\begin{equation} 
    \begin{split}
        -\frac{n\lambda  r^2}{2} + \frac{1}{\theta}\sum_{i=1}^n x_i & = 0 \\
        \frac{1}{\theta}\sum_{i=1}^n x_i    & = \frac{n\lambda  r^2}{2} \\
        2\cdot\frac{\sum_{i=1}^n x_i}{n\lambda r^2}    & = \theta
    \end{split}
\end{equation}

\begin{figure}
    \centering
    \includegraphics[width=0.95\linewidth]{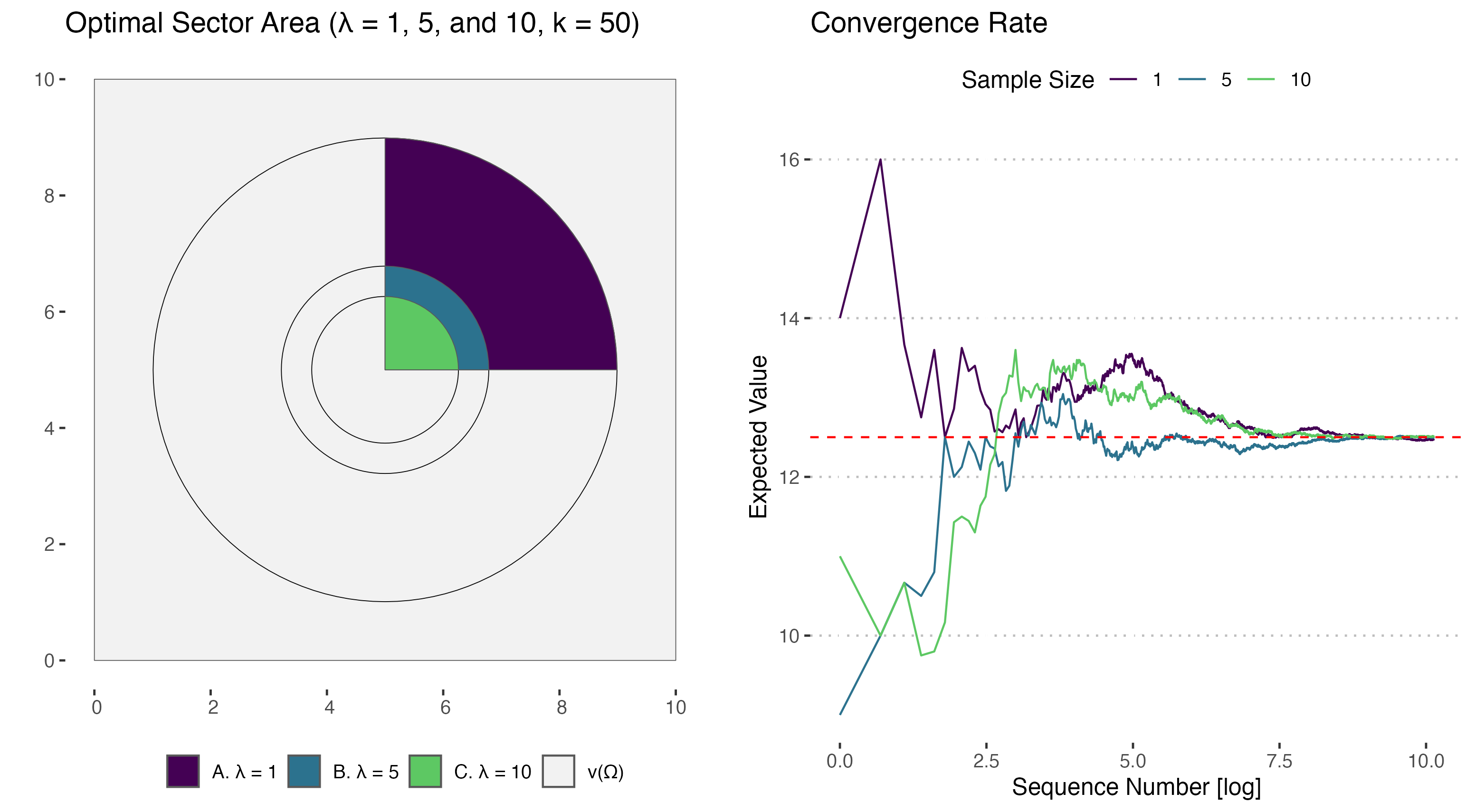}
    \caption{Convergence Rate for Optimal Radius: Stationary Point Process (k = 50)}
    \label{fig:convergenceCentral}
\end{figure}

The left graph of Figure \ref{fig:convergenceCentral} visualizes these restricted geofence perimeters while the right graph illustrates how they converge towards the new privacy constraint $k = 12.5$. 

%% file: 3_Appendicies/Appendix_Polygon.tex
\section{Appendix 8: Polygon Adaptive Plug-In Estimator}
\label{appx:polygon}
We derive the polygon adaptive estimator by substituting $\nu(B(a_i,r))$ in the focal adaptive estimator with the area formula for a regular sided polygon $A$ (see Equation \ref{eq:e1}). We use the area formula based on the n-sided polygon's circumradius.  We derive the circumradius plug-in estimator with the following steps. Equation \ref{eq:e2} describes the modified log-likelihood function.  We derive the partial derivatives for each of the three parameters $(\lambda,r,p)$. We also provide optimal radius estimator based on the apothem and side length of a regular sided polygon. 

\begin{equation}
        A=\frac{p\cdot\\r^2}{2}\sin (\frac{2\pi }{p})
        \label{eq:e1}
\end{equation}

\begin{equation}
    \mathcal{LL}(\lambda,r,p) = -\frac{n \cdot\lambda \cdot p \cdot r^2 \cdot\sin(\frac{2\pi}{p})}{2} - \sum_{i=1}^{n}ln(x_i!) + ln(\frac{\lambda \cdot p \cdot r^2 \cdot \sin(\frac{2\pi}{p})}{2}) \sum_{i=1}^n x_i
    \label{eq:e2}
\end{equation}

\begin{definition}[Circumradius Plug-In Estimator]
    \begin{equation}
        \hat{r}_{\text{Circumradius}} = \sqrt{2\cdot\frac{k}{\lambda \cdot p \cdot \sin(\frac{2\pi}{p})}} 
    \end{equation}
\end{definition}

\begin{definition}[Apothem of Inscribed Polygon]
    \begin{equation}
        \hat{r}_{\text{Apothem}} = \hat{r}_{\text{Circumradius}} \cos (\frac{\pi}{p}) = \sqrt{2\cdot\frac{k}{\lambda \cdot p  \cdot\sin(\frac{2\pi}{p})}} \cos (\frac{\pi}{p})
    \end{equation}
\end{definition}

\begin{definition}[Side Length of Polygon]
    \begin{equation}
        L=2\sin (\frac{\pi }{p})\hat{r}_{\text{Circumradius}} = 2\sin (\frac{\pi }{p})\sqrt{2\cdot\frac{k}{\lambda \cdot p \cdot \sin(\frac{2\pi}{p})}}
    \end{equation}
\end{definition} 

\paragraph{Partial-Derivative for $\lambda$:}
\begin{equation} 
    \begin{split}
        \frac{\partial}{\partial\lambda}\mathcal{LL(\lambda,r)} & =  \frac{\partial}{\partial\lambda} \Biggr[ -\frac{n\lambda vr^2 \sin(\frac{2\pi}{v})}{2} - \sum_{i=1}^{n}ln(x_i!) + ln(\frac{\lambda vr^2 \sin(\frac{2\pi}{v})}{2}) \sum_{i=1}^n x_i \Biggr] \\
        & = -\frac{nvr^2\sin(\frac{2\pi}{v})}{2} + \frac{2vr^2\sin(\frac{2\pi}{v})}{2\lambda vr^2 \sin(\frac{2\pi}{v})}\sum_{i=1}^n x_i \\
        & = -\frac{nvr^2\sin(\frac{2\pi}{v})}{2} + \frac{1}{\lambda}\sum_{i=1}^n x_i
    \end{split}
\end{equation}

\begin{equation} 
    \begin{split}
        -\frac{nvr^2\sin(\frac{2\pi}{v})}{2} + \frac{1}{\lambda}\sum_{i=1}^n x_i  & = 0 \\
        \frac{\sum_{ i=1}^n x_i}{\lambda} & = \frac{nvr^2\sin(\frac{2\pi}{v})}{2}  \\
        \sum_{i=1}^n x_i & = \frac{n\lambda vr^2\sin(\frac{2\pi}{v})}{2} \\
        2\cdot\frac{\sum_{i=1}^n x_i}{n vr^2\sin(\frac{2\pi}{v})} & = \lambda  \\
    \end{split}
\end{equation}

\paragraph{Partial-Derivative for $r$:}
\begin{equation} 
    \begin{split}
        \frac{\partial}{\partial r}\mathcal{LL(\lambda,r,\theta)} & =  \frac{\partial}{\partial r} \Biggr[ -\frac{n \cdot\lambda \cdot p \cdot r^2 \cdot\sin(\frac{2\pi}{v})}{2} - \sum_{i=1}^{n}ln(x_i!) + ln(\frac{\lambda \cdot p \cdot r^2 \cdot \sin(\frac{2\pi}{p})}{2}) \sum_{i=1}^n x_i \Biggr]  \\
        & = -\frac{2n \cdot\lambda \cdot p \cdot  r \cdot\sin(\frac{2\pi}{p})}{2} + \frac{4 \lambda \cdot p \cdot  r \cdot\sin(\frac{2\pi}{p})}{2\lambda \cdot p \cdot r^2 \cdot \sin(\frac{2\pi}{p})}\sum_{i=1}^n x_i \\
        & = -n\lambda \cdot p \cdot  r \cdot\sin(\frac{2\pi}{p}) + \frac{2}{r}\sum_{i=1}^n x_i
    \end{split}
\end{equation}

\begin{equation} 
    \begin{split}
        -n \cdot\lambda \cdot p \cdot  r \cdot\sin(\frac{2\pi}{p}) + \frac{2}{r}\sum_{i=1}^n x_i & = 0 \\
        \frac{2}{r}\sum_{i=1}^n x_i    & = n \cdot\lambda \cdot p \cdot  r \cdot\sin(\frac{2\pi}{p}) \\
        2\sum_{i=1}^n x_i    & = n \cdot\lambda \cdot p \cdot  r^2 \cdot\sin(\frac{2\pi}{p}) \\
        \sqrt{2\cdot\frac{\sum_{i=1}^n x_i}{n \cdot\lambda \cdot p \cdot \sin(\frac{2\pi}{p})}}    & = r \rightarrow \sqrt{2\cdot\frac{k}{\lambda \cdot p \cdot \sin(\frac{2\pi}{p})}}    = r 
    \end{split}
\end{equation}

\begin{figure}
    \centering
    \includegraphics[width=0.85\linewidth]{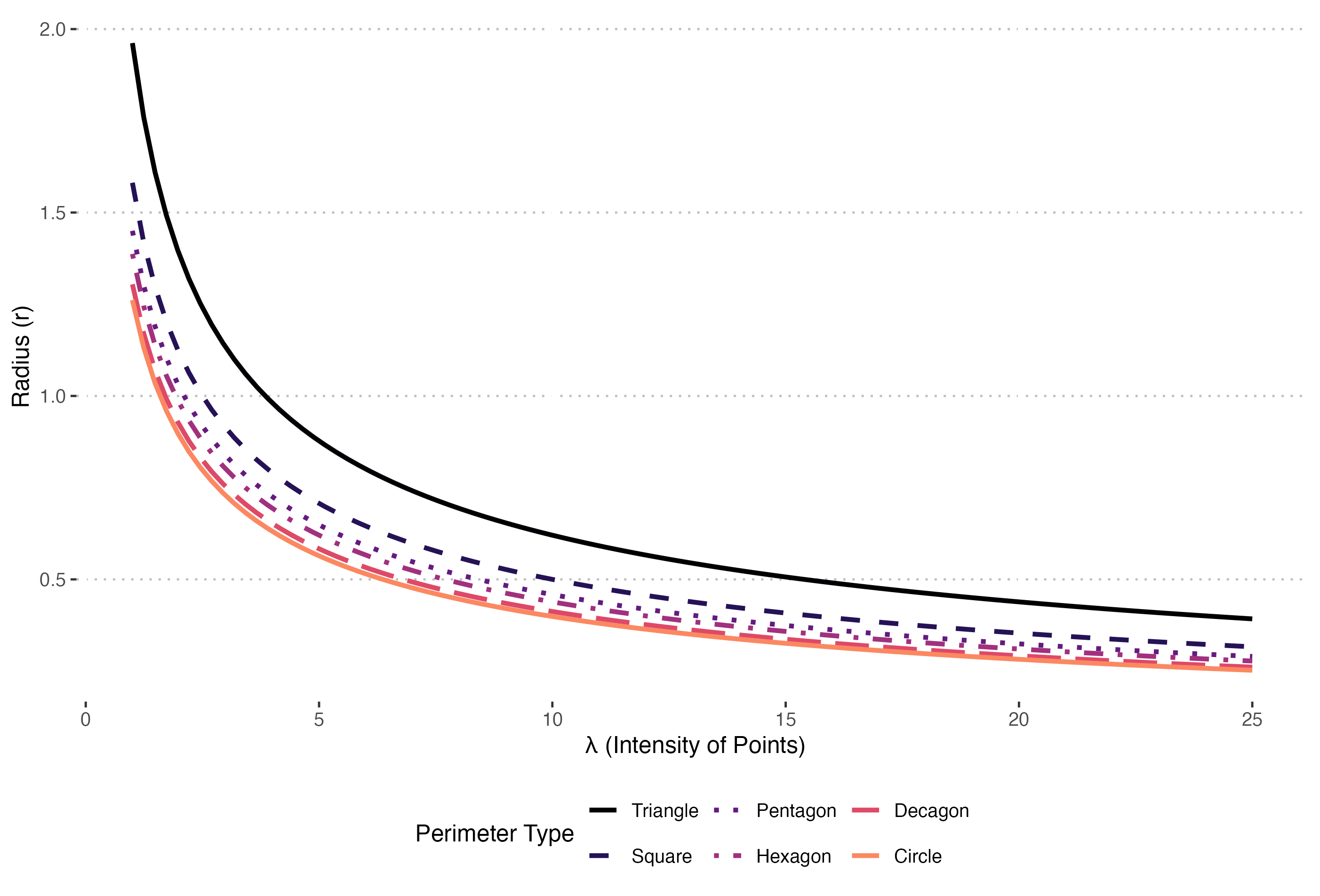}
    \caption{Polygon Circumradius Size and Point Intensity (k = 5)}
\end{figure}

\begin{figure}
    \centering
    \includegraphics[width=0.95\linewidth]{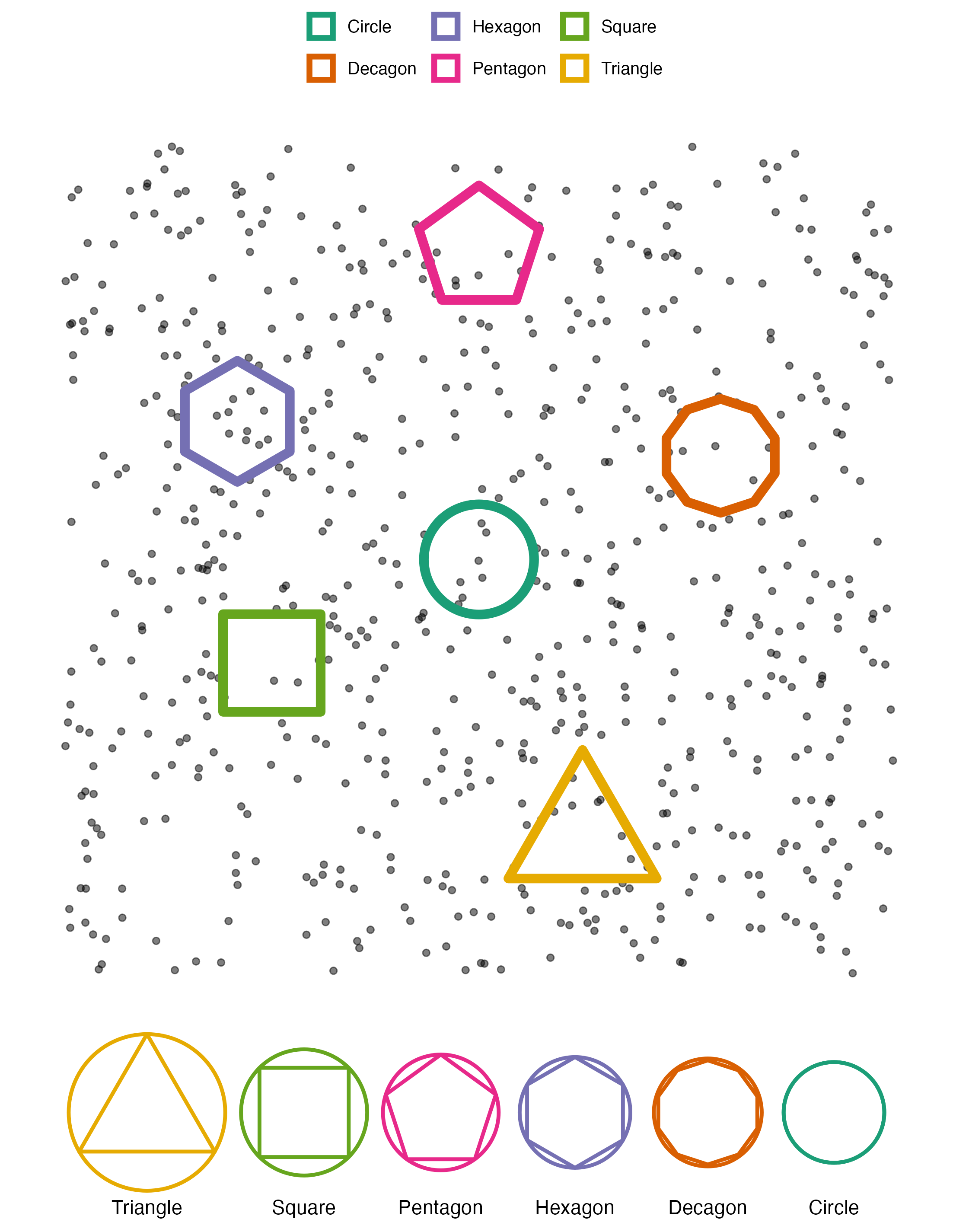}
    \caption{Visual Comparison of Polygon Perimeters (k = 10)}
    \label{fig:polygon}
\end{figure}

Figure \ref{fig:polygon} visually compares several regular $p$-sided polygon perimeters with a circular perimeter. Each perimeter returns, on average, $k=10$ geolocated individuals; assuming that individuals are uniformly distirbuted across the jurisdiction.

%% file: 3_Appendicies/Appendix_Simulation.tex
\section{Appendix 9: Technical Details for Agent-Based Simulation}
\label{appx:simulation}
Our agent-based simulation attempts to mimic the data generating process for real-world datasets on cell phone geolocation. We use agent-based random walks to simulate migratory pathways. Each walk generates a set of geolocation tags that we then use to generate our privacy measures. We define these random walks with the following notation. First, let $S_{{n_s} \times {m_i}}$ represent the random walk for agent $m_i \in M$ after $n_s$ steps, where $n_s \in \mathbbm{N}$, $S_0 \in \mathbbm{R}^2$ and $S_{n_s} = S_0 + \sum_{i=1}^{n_s} X_i$. Next, let $\eta(\{S_{n_s \times m_i}\})$ represent the collection of geolocation tags agent $m_i$ produces after $n_s$ steps. 

The shifting spatial distribution of these agents across multiple step periods generates the necessary variation that we need to compare our geofence perimeters. We vary these random walks range from unbiased - agents neither interact with each other nor with targets inside the window - to endogenously biased - agents adapt their random walks by interacting (clustering or repelling) with each other and fixed targets. We relied on Gemini 3.1 pro to help us write code for our agent-based model. Table \ref{tab:simParameters} summarizes the parameters we used to vary the complexity of our agent's random walks. 

\begin{table}[]
\centering
\resizebox{\columnwidth}{!}{%
\begin{tabular}{|p{4cm}|p{7cm}|p{7cm}|}
\hline
\multicolumn{1}{|c|}{\textbf{Category}} &
  \multicolumn{1}{c|}{\textbf{Description}} &
  \multicolumn{1}{c|}{\textbf{Distribution}} \\ \hline
Number of Steps &
  How many steps agents take during the simulation round &
  Uniform(1,1000) \\ \hline
Agent Starting Positions &
  Longitude/latitude coordinates of agent's starting positions &
  Poisson($\Lambda(\Omega)$) or exp($Z(\Omega)$). $Z$ is the gaussian random field GauPro(\textbf{$\mu$}, $C_{\nu}(d)$), where $C_{\nu}(d)$ is the Mat\'ern covariance function
  \begin{itemize}
    \item Lambda ($\lambda$) (Random Starts): Uniform(0.01, 1)
      \item Psill: Uniform(0.25, 2.5)
      \item Range: Uniform(2, $\sqrt{X^2 + Y^2}$)
      \item Nugget: Uniform(0, 0.25)
      \item Beta ($\mu$): Uniform(-2.5, 1)
  \end{itemize}\\ \hline
Fixed Target Position &
  Longitude/latitude coordinates of fixed targets &
  Uniform($\Omega$) \\ \hline
Interaction Radius &
  Radius that determines whether agents interact with each other or not.  Agents whose distance are less than this interaction radius will interact.  Agents whose distance are more than the interaction radius act independently. &
  Uniform(0,5) \\ \hline
Agent Interaction Strength &
  Strength of clustering or repulsion tendencies between agents whose distance are less than the interaction radius &
  Uniform(-1,1) \\ \hline
Target Interaction Strength &
  Strength of clustering or repulsion tendencies between agents and fixed targets. &
  Uniform(-1,1) \\ \hline
Pause Probability &
  Probability that agent will pause their random walk &
  Uniform(0,0.5) \\ \hline
Maximum Number of Paused Steps &
  The number of step periods an agent will wait until they resume their random walk &
  Poisson(\# of Steps $\cdot$ 0.25) \\ \hline
X- and Y-Range Maximum &
  Maximum value for x- and y-ranges of the rectangular study window. Minimum values are both set equal to zero. &
  Uniform(10,35) \\ \hline
\end{tabular}%
}
\caption{}
\label{tab:simParameters}
\end{table}

We prepared each simulation round with the following steps. First, we randomly generated a study window with variable height and width. Second, we randomly sampled 30 surveillance sites inside the study window. Third, for each surveillance site, we assigned the following geofence perimeters: fixed, window, focal (adaptive and fixed), and lambda (adaptive and fixed). We will use $D$ to reference this set of geofence perimeters. Fourth, we randomly sampled a privacy constraint $k$ and adjusted all geofence perimeters (excluding fixed perimeters) accordingly. Fifth, we sampled agent starting locations from either a stationary Poisson process (Poisson($\Lambda(\Omega)$) or an exponentiated Gaussian random field (exp($Z(\Omega)$)). We use the Mat\'ern covariance function $C_{\nu}(d)$ to define the Gaussian process' covariance. Table \ref{tab:simParameters} summarizes the uniform distributions we sample for each parameter. For the range parameter, we sample values on the uniform interval between 2 and $\sqrt{X^2 + Y^2}$, which is the hypotenuse of the study window. 

During each step period, we adjust the boundaries of adaptive perimeters according to their relevant parameters. For fixed perimeters, we randomly sample the sampling region's radius at the beginning of the simulation round. We fix this size across all step rounds. For window perimeters, we adjust their bounds according to the number of agents inside the study window boundaries during each step period. For focal perimeters, we adjust their bounds according to the density of agents at the surveillance site. For lambda perimeters, we begin with the focal perimeter estimate and then adjust perimeter size according to the density of agents inside its convex hull during each step period. 

We adopt the following steps to calculate our first outcome measure for each geofence. We begin with the random summation in Equation \ref{eq:agentCount} to calculate the total number of intersecting agents $m_i \in M$ for the geofence $j$ during the step period $k \in S$: $Y_{M \times D_j \times s_k}$. Here, $D_j$ represents  the geofence $j$, $m_i \in M$ represents an agent from the simulation round, and $s_k$ represents the step period $k$. $\mathbbm{1}_{D_j}$ represents the binary indicator variable that returns a value of one if the agent $m_i$ is inside the geofence $D_j$ and zero otherwise. $\mathbbm{1}_{s_k}$ represents the binary indicator variable that returns a value of one for the step period $k$. Next, we use Equation \ref{eq:AbsoluteDev} to calculate the absolute deviation score for the geofence $j$ during the step period $k$. Finally, we use Equation \ref{eq:meanAbsoluteDev} to calculate the mean absolute deviation score for the geofence $j$ across all step periods $S$. 

\begin{equation}
    Y_{D_j \times M \times s_k} = \sum_{m_i \in M} \bigg [\mathbbm{1}_{D_j}\mathbbm{1}_{s_k} \bigg ]
    \label{eq:agentCount}
\end{equation}

\begin{equation}
   \delta_{jk} = |Y_{D_j \times M \times s_k} - k|
   \label{eq:AbsoluteDev}
\end{equation}

\begin{equation}
    \bar{\delta_j} = \frac{1}{|S|}\sum_{k=1}^{|S|} \delta_{jk}
    \label{eq:meanAbsoluteDev}
\end{equation}

We adopt the following steps to calculate our second outcome measure for each agent. We begin with the random summation in Equation \ref{eq:stepCount} to calculate the total number of steps (geolocation tags) $s_k \in S$ that intersect with the geofence $j \in D$ for the agent $i$: $M_{i \times S \times D_j}$. 

\begin{equation}
    M_{i \times S \times D_j} = \sum_{s_k \in S} \bigg [\mathbbm{1}_{D_j}\bigg ]
    \label{eq:stepCount}
\end{equation}

\subsection*{Proof A: Proof for Expected Time Under Surveillance}
\label{proofA_sim}
\begin{proof}
    In this proof, we will prove that the expected time under surveillance equals $k/n$ when agents are randomly distributed across all step rounds. We will define time under surveillance as the expected number of steps inside the perimeter divided by the total number of steps an agent takes (Definition 1). Let $n$ agents be randomly distributed inside the study window $\Omega$ across $s$ steps. The intensity of agents equals $n/\nu(\Omega)$ and remains constant across the entire study window and all step rounds. 
    
    Let $B$ represent a geofence perimeter that captures $k$ individuals, on average. We can use the focal adaptive plug-in estimator $\hat{r}_{Pois}$ to estimate the total area for $B$, where the area equals $\pi\cdot \hat{r}_{Pois}^2$. We must swap $\lambda(a_i)$ with our constant intensity $n/\nu(\Omega)$ since the intensity is constant across the entire study window. This gives us the following formula for its area:

    \begin{equation}
        \pi \cdot \bigg ( \sqrt{\frac{k}{\pi \cdot \big (\frac{n}{\nu(\Omega)} \big )}} \bigg )^2
    \end{equation}

    \noindent This equation further reduces to $\frac{k \cdot \nu(\Omega)}{n}$.

    \begin{equation}
        \pi \cdot \bigg ( \sqrt{\frac{k}{\pi \cdot \big (\frac{n}{\nu(\Omega)} \big )}} \bigg )^2 = \pi \cdot \frac{k}{\pi \cdot \big (\frac{n}{\nu(\Omega)} \big )} = \frac{k \cdot \pi \cdot \nu(\Omega)}{\pi \cdot n} = \frac{k \cdot \nu(\Omega)}{n}
    \end{equation}
    
    \noindent For a single step round, the probability that an agent is captured by $B$ equals the ratio between the sizes of $B$ and $\Omega$. This gives us the following formula for its probability:

    \begin{equation}
        p =\frac{\frac{k \cdot \nu(\Omega)}{n}}{\nu(\Omega)} = \frac{k \cdot \nu(\Omega)}{n \cdot \nu(\Omega)} = \frac{k}{n}
    \end{equation}

    \noindent For $s$ step rounds, we can expect, on average, that the agent will be inside the perimeter for $s\cdot\frac{k}{n}$ steps. Given Definition 1, the expected "time under surveillance" for agents randomly distributed across the study window and all step periods equals $k/n$.
\end{proof}

%% file: 3_Appendicies/Appendix_Average_Time.tex
\section*{Appendix 10: Results for Average Time Under Surveillance}
\label{appx:average_time}
How much does an agent's observed time under surveillance differ from their expected time for surveillance? We calculate the observed time under surveillance as the number of steps taken inside the geofence divided by the total number of steps taken during a simulation round. As a reminder, our baseline expectation involves a scenario where agents randomly distribute their geolocation tags across the study window for all time periods. When this occurs, each agent can expect approximately $s\cdot p$ of their steps to intersect with a geofence $B$, where $s$ is the number of steps and $p$ is the proportion of area covered by the geofence inside the study window. In this baseline expectation, we assume that the geofence $B$ perfectly captures $k$ individuals. This implies that the expected time for surveillance equals $k/n$ (see Proof A in Appendix 9). In other words the expected time for surveillance equals the proportion of  individuals that would be captured if the total population was uniformly distributed. 

Table \ref{table:errorSurveillance} summarizes partial dependence estimates for each perimeter along with bootstrapped standard errors and confidence intervals. Similar to Table \ref{table:deviance}, the perimeters derived from our proposed estimators perform better. Window adaptive perimeters saw approximately a 34 percent reduction in its mean absolute deviation. Focal adaptive perimeters saw approximately a 45 percent reduction. Finally, lambda adaptive perimeters saw approximately a 95 percent reduction. These reductions were similar regardless of whether agents had random starting positions or not. 
 
\begin{table}[hbt!]
\begin{center}
\begin{threeparttable}%
\begin{tabular}{llllll}
\toprule
Perimeter Type & Random Start & Mean & SE & q0.025 & q0.975 \\ 
\toprule
Fixed & No & 0.361 & 0.0091 & 0.340 & 0.375 \\ 
 & Yes & 0.369 & 0.0091 & 0.350 & 0.384 \\ 
  \midrule
Window & No & 0.239 & 0.0071 & 0.226 & 0.252 \\ 
 & Yes & 0.237 & 0.0072 & 0.225 & 0.249 \\ 
  \midrule
Focal & No & 0.199 & 0.0050 & 0.189 & 0.208 \\ 
 & Yes & 0.197 & 0.0053 & 0.189 & 0.207 \\ 
  \midrule
Focal (Fixed) & No & 0.212 & 0.0052 & 0.203 & 0.222 \\ 
 & Yes & 0.211 & 0.0056 & 0.200 & 0.222 \\ 
  \midrule
Lambda & No & 0.019 & 0.0009 & 0.018 & 0.021 \\ 
 & Yes & 0.020 & 0.0009 & 0.018 & 0.021 \\ 
  \midrule
Lambda (Fixed) & No & 0.021 & 0.0010 & 0.019 & 0.023 \\ 
 & Yes & 0.021 & 0.0010 & 0.020 & 0.023 \\ 
\bottomrule
\end{tabular} 
\end{threeparttable}
\end{center}
\caption{}
\label{table:errorSurveillance}
\end{table} 

\begin{figure}[h]
    \centering
    \includegraphics[height = 3.5in, keepaspectratio]{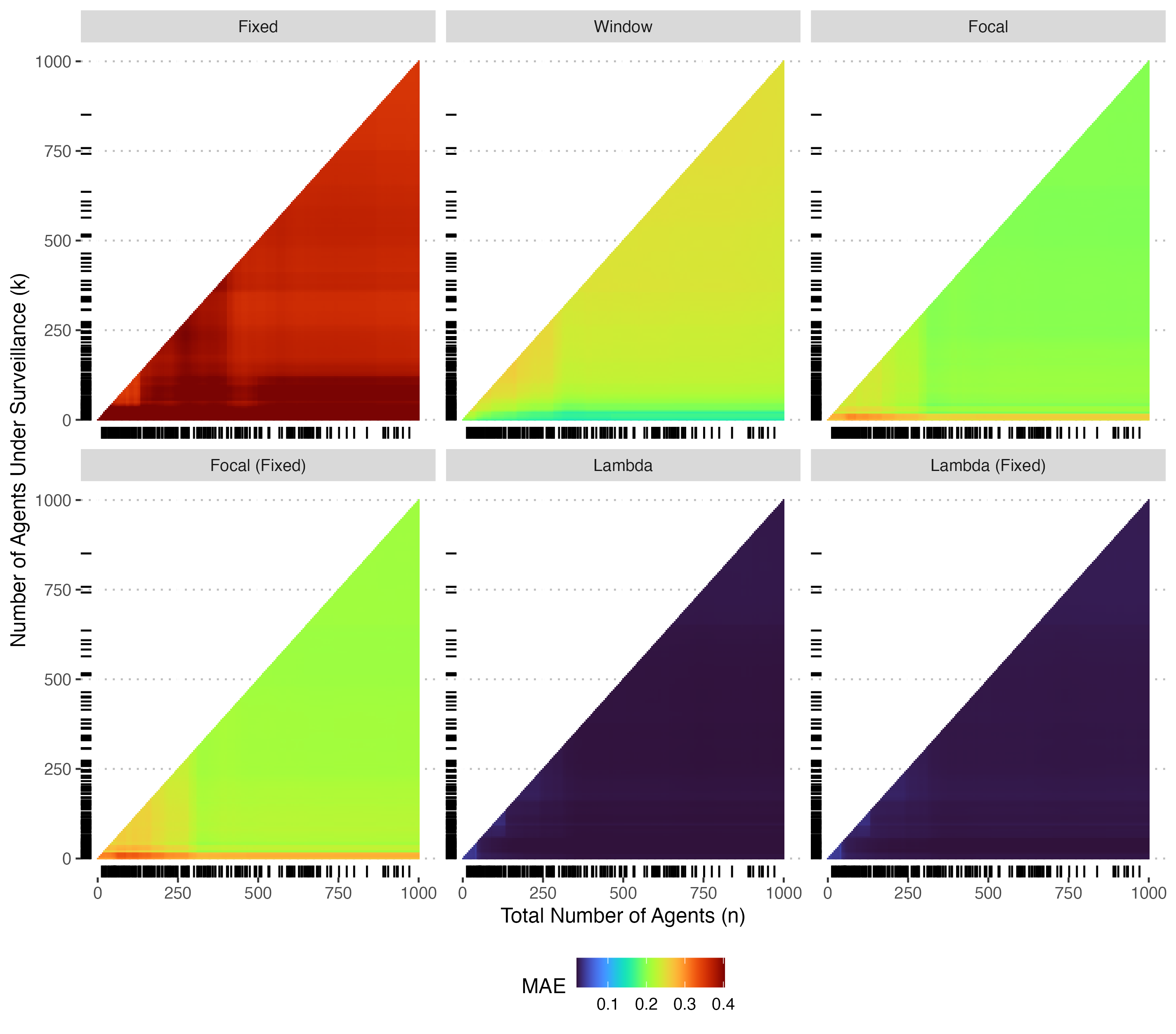}
    \caption{Bivariate Partial Dependence Plot (Deviation)}
    \label{fig:bivariateError}
\end{figure}

\begin{figure}[h]
\centering
\begin{minipage}{.5\textwidth}
  \centering
    \includegraphics[height = 2.5in, keepaspectratio]{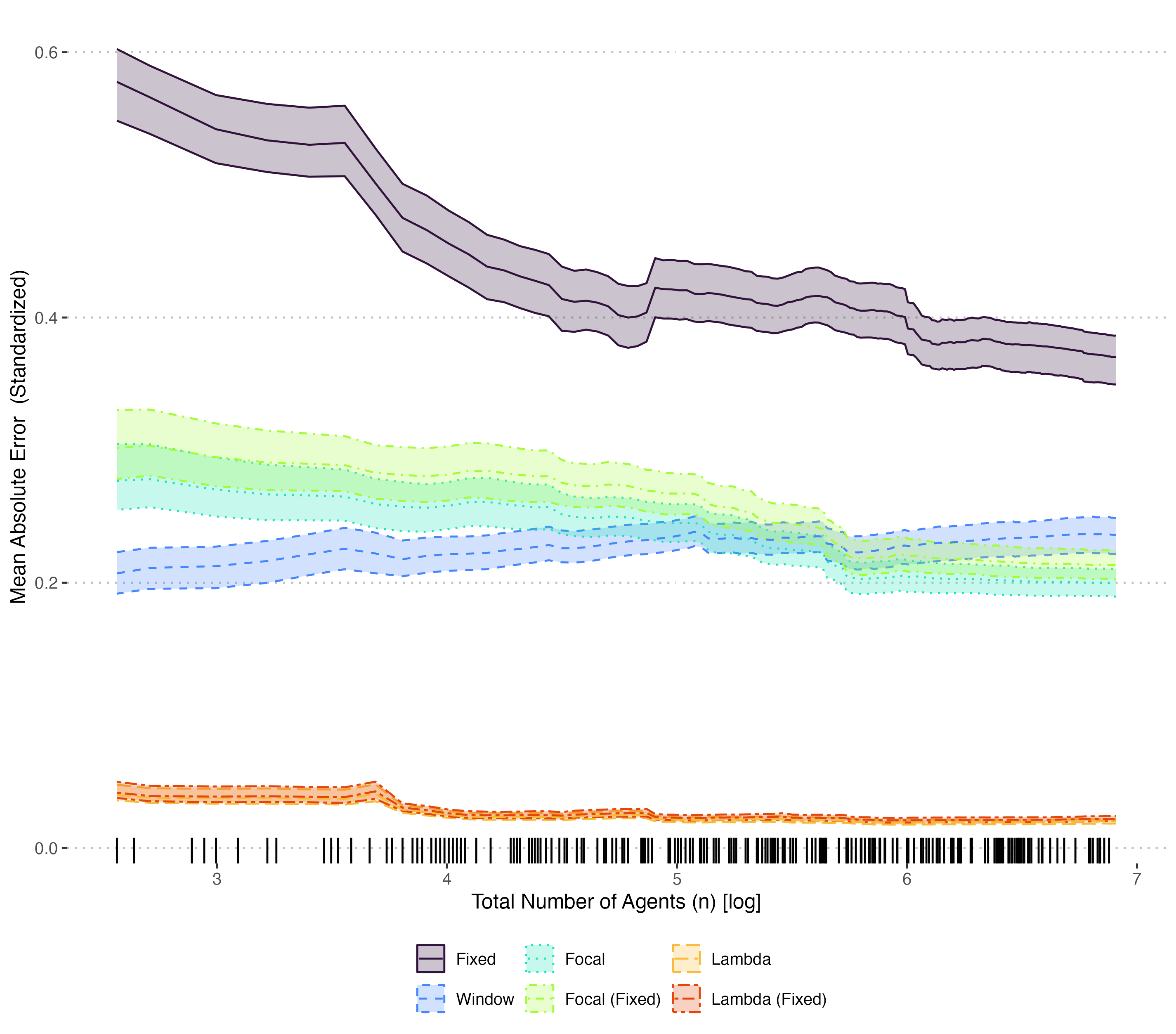}
    \caption{Total Number of Agents ($n$)}
    \label{fig:marginalError_total}
\end{minipage}%
\begin{minipage}{.5\textwidth}
  \centering
    \includegraphics[height = 2.5in, keepaspectratio]{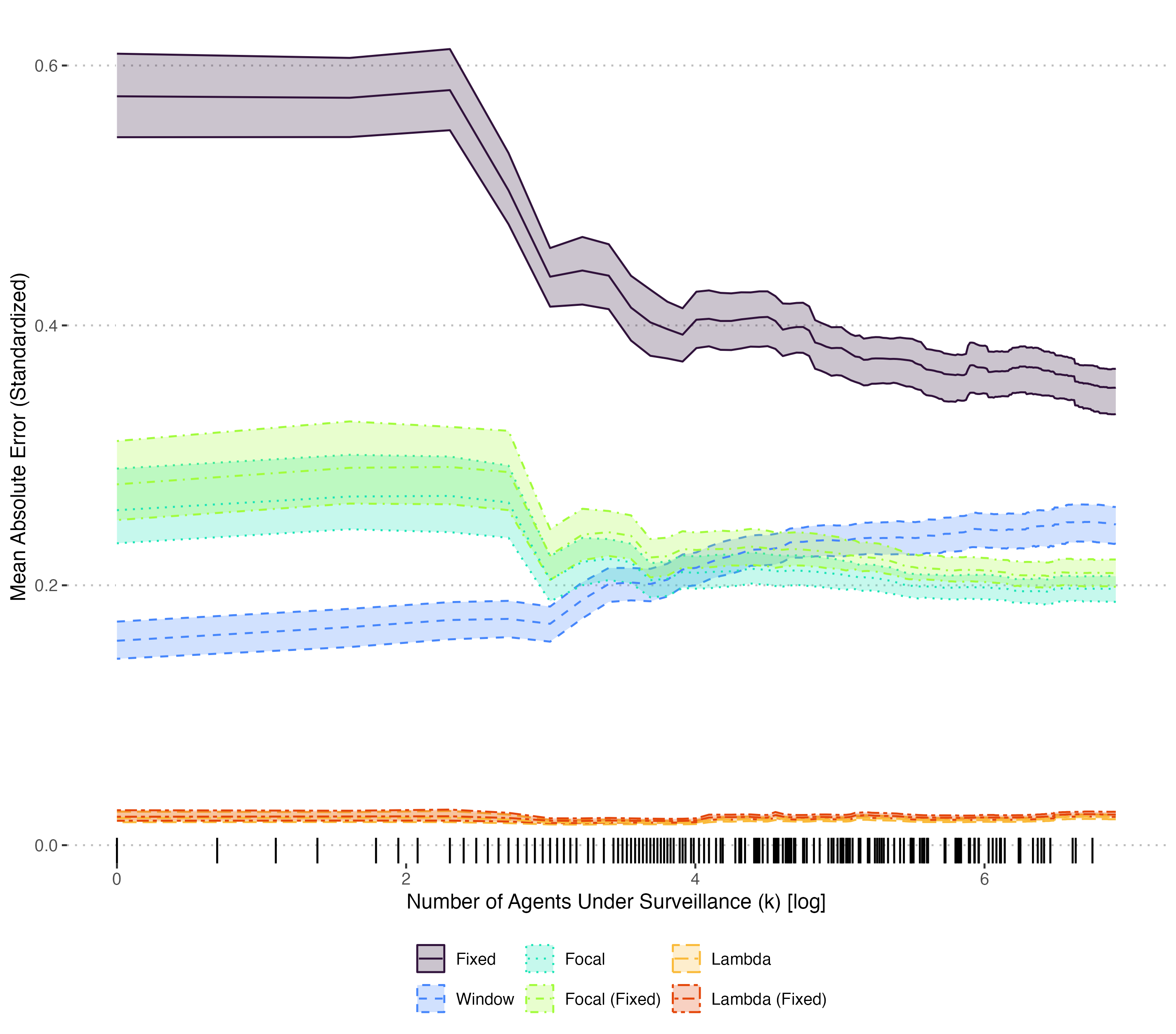}
    \caption{Privacy Constraint ($k$)}
    \label{fig:marginalError_privacy}
\end{minipage}
\end{figure}

The partial dependence plot in Figure \ref{fig:bivariateError} shows a clear trend. The estimator-derived perimeters performed better at all combinations of $k$ and $n$ than the fixed perimeter. Figures \ref{fig:marginalError_total} and \ref{fig:marginalError_privacy} replicate these trends at the margins. In both figures, we log the x-axes to better visualize the trends since smaller values tend to have larger errors. In general, performance improves as the total number of agents and privacy constraints increase. Window perimeters being the exception, since it tends to under surveil agents relative to the the expected baseline $k/n$. 

